\documentclass[amsmath,amssymb,onecolumn,superscriptaddress]{revtex4}

\usepackage{graphicx}%
\usepackage{dcolumn}%
\def\be{\begin{equation}}
\def\ee{\end{equation}}
\def\beq{\begin{eqnarray}}
\def\eeq{\end{eqnarray}}

\usepackage{bm}
\usepackage{graphicx}
\usepackage{dcolumn}
\usepackage{amsmath}
\usepackage[latin1]{inputenc}
\usepackage{graphicx, psfrag}
\usepackage{amssymb}
\usepackage[colorlinks=true, citecolor=blue, urlcolor = blue, linkcolor= red, bookmarks=true]{hyperref}
\usepackage{float}
\usepackage{mathtools}
\usepackage{amsmath}
\usepackage{amsfonts}
\usepackage{dcolumn}
\usepackage{hyperref}
\usepackage{subfigure}
\usepackage{pgfplots}
\usepackage{epstopdf}
\usepackage{booktabs}
\usepackage{orcidlink}
\usepackage{xcolor}

\begin{document}
\title{Polytropic stellar structure in 5$\mathcal{D}$ Einstein-Gauss-Bonnet gravity}

\author{Akashdip Karmakar \orcidlink{0009-0007-3848-1443}}
\email[Email:]{akashdip999@gmail.com}
 \affiliation{Department of Mathematics,  Indian Institute of Engineering Science and Technology, Shibpur, Howrah, West Bengal 711 103, India.}

\author{Ujjal Debnath}
\email[Email:]{ujjaldebnath@gmail.com}
 \affiliation{Department of Mathematics,  Indian Institute of Engineering Science and Technology, Shibpur, Howrah, West Bengal 711 103, India.}
 
\author{Pramit Rej \footnote{Corresponding author} \orcidlink{0000-0001-5359-0655}}
\email[Email:]{pramitrej@gmail.com, pramitr@sccollegednk.ac.in, pramit.rej@associates.iucaa.in}
 \affiliation{Department of Mathematics, Sarat Centenary College, Dhaniakhali, Hooghly, West Bengal 712 302, India.}

\begin{abstract}
Polytropic stars are useful tools for learning about stellar structure without the complexity of comprehensive stellar models. These models rely on a certain power-law correlation between the star's pressure and density. This paper proposes a polytropic star model to investigate some new features in the context of $5\mathcal{D}$ Einstein-Gauss-Bonnet (EGB) gravity using the Finch-Skea {\em ansatz} [{\em M. R. Finch and J. E. Skea, Classical and Quantum Gravity 6, 467 (1989)}]. Analytical results are better described by graphical representations of the physical parameters for various values of the Gauss-Bonnet coupling constant $\alpha$. The solution for a specific compact object, EXO 1785-248, with radius $\mathfrak{R} = 8.849_{-0.04}^{+0.04}$ km and mass $\mathcal{M} = 1.3 \pm 0.02~\mathcal{M}_{\odot}$, is shown here. We analyze the essential physical attributes of the star, which reveal the influence of the coupling parameter $\alpha$ on the values of the parameters. Ultimately, we conclude that our current model is realistic because it satisfies all the physical criteria for an acceptable model. 
\end{abstract}

\maketitle

\maketitle

\section{Introduction}
The most notable development in contemporary science is the theory of general relativity (GR). Our knowledge of astrophysical compact objects (CO) has improved as a result of this theory. Theoretical studies and modeling have advanced greatly in response to the difficulties of obtaining exact analytic solutions of Einstein field equations characterizing compact structures. 
Despite its wide acceptance, Einstein's general theory of relativity has many shortcomings, such as its inability to explain the accelerating expansion of the universe. Furthermore, beyond the 4-dimensional context, the aforementioned theory does not reflect successfully. Two different strategies have been implemented to address these problems.
One is to alter the gravitational component of the Einstein-Hilbert action, and the other is to change the matter component of Einstein's theory, which results in the dark matter and dark energy hypotheses. Many modified theories of gravity have surfaced, including the second one. An ideal illustration of how to explain the distinctions between General Relativity (GR) and its modifications is the study of a massive gravitational field in dense compact objects. Several modified theories of gravity spanning a wide range of issues have been documented in the literature \cite{shahzad2019strange, bhatti2021dynamical, goswami2014collapsing, hansraj2022strange, bhar2022tolman, rej2023isotropic, rahaman2020anisotropic, salti2017brans}. In this connection, it is to be noted that a lot of research has been done on the effects of local anisotropy on the global properties of relativistic compact objects as anisotropic matter distributions, or unequal radial and transverse pressure ($p_r \neq p_t$), are crucial to the stability and equilibrium of the stellar structure \cite{maurya2019generalized, tamta2017new, dev2002anisotropic, krori1984some, maurya2016new}. In this particular instance, the analysis of static configurations with spherical symmetry composed of isotropic pressure distributions and ideal fluid distributions (i.e., $p_r = p_t$) is the most straightforward scenario.\\
Higher-dimensional theories of gravity have sparked a great deal of attention for almost half a century. Sometimes, imagining other dimensions helps to make sense of the unexplained events associated with gravity \cite{arkani1998hierarchy, randall1999alternative}. In the quest for a self-consistent gravity theory, braneworlds and other higher-dimensional adaptations of Einstein's general relativity, such as Lovelock theory, were regarded as tenable extensions. Indeed, Lovelock gravity has been proposed as one of the higher-dimensional gravity theories with the idea that higher-order corrections to Einstein's theory may resolve the singularity problem with black holes (BHs), avoiding classical causality concerns, for instance \cite{Lovelock1972TheFO}. Notably, Lovelock theory takes general relativity further into higher-dimensional spacetimes while maintaining the order of the field equations at second order in derivatives without torsion. The GR and the modified gravity theories might explain the gravitational effects with regard to matter configuration and spacetime curvature. Torsion can be used in place of curvature to create and illustrate a similar theory of GR. Since the entire Riemann curvature tensor may be assumed to be zero in this situation, torsion can be utilized to describe the gravitational field. This leads to the teleparallel equivalent of GR (TEGR), an alternative explanation of GR \cite{abbas2015strange, moller1961further}. A common goal of scientific research on modified theories of gravity is to identify a change in the Cosmological Constant as the reason behind the current rapid expansion of the Universe. Nevertheless, it may also be used to investigate and, as a result, limit any theory that might arise from high-energy adjustments to GR, such as string theory \cite{karmakar2023charged}. The standard formulations of string theory need a total of ten dimensions, or eleven if we consider a modified version referred to as $M-$Theory, which postulates the presence of extra spacetime dimensions in addition to the four \cite{kaluza2018unification}. In contrast to earlier research conducted in four dimensions ($4\mathcal{D}$), cosmic censorship in higher dimensions has yielded intriguing results. Gravity cannot be studied in fewer than four dimensions. In terms of two-dimensional spacetime, the Euler property of the Einstein-Hilbert Lagrangian has the effect of making the Einstein tensor cease to exist. Although an Einstein tensor that endures in three dimensions already exists, gravitational wave solutions are not included in the Ricci-plane solutions (or those derived from $R_{\mu\nu} =0$), which are the outcomes of the vanishing Riemann tensor. After all, we have the typical three spatial dimensions plus a time dimension. So, based on the previous argument, it may be fair to assume that further research in more extended dimensions will prove beneficial. Thus, there is no requirement to avoid terms of scalars that are quadratic, cubic, etc., that are derived from the Riemann tensor and its contractions in higher dimensions \cite{bhar2019compact}.
The most basic non-trivial Lovelock gravity among all of the categories is the EGB gravity, which features a Lagrangian that is the product of a curvature scalar and a cosmological constant, together with a quadratic Gauss-Bonnet (GB) term in the third term. In a $\mathcal{N}$-dimensional spacetime (where $\mathcal{N} \geq 5$), the GB Lagrangian produces the necessary second-order equations of motion. A helpful generalization of classical general relativity is EGB gravity, which is produced by adding a term to the fundamental Einstein-Hilbert action and is quadratic in the Riemann tensor. Stability in $5\mathcal{D}$ EGB can be achieved with a greater mass than in traditional $4\mathcal{D}$ Einstein gravity, which is another fascinating result. Although the EGB does hint at discoveries on the significance of gravitational collapse, it is important to remember that the GR and the EGB are equivalent in the $4\mathcal{D}$ scenario. Several models have been examined in the literature in higher dimensions in addition to the conventional $4\mathcal{D}$ framework, which makes use of the EGB theory \cite{hansraj2015exact, boulware1985string, myers1986black, bhar2017comparative}.\\
The premise of a polytropic equation of state (EoS), i.e., $p_r = \gamma \rho^{1 + \frac{1}{\eta}} + \beta\rho + \chi$, forms the basis of our investigation. We determine our solution for a polytropic EoS with a specified polytropic index, $\eta = 1$. The polytropic EoS, $p_r = \gamma \rho^{1 + \frac{1}{\eta}}$ has been extensively utilized in the literature to examine the characteristics of compact objects \cite{herrera2016cracking, takisa2013some, azam2017cracking, cosenza1981some}.
 Because polytropes are self-gravitating gaseous spheres, they can be used as a rough approximation to more accurate stellar models. Furthermore, it helps to describe the internal structure of neutron stars (NSs), including their maximum mass, surface temperature, pulsar glitches, and other features, because it fits the EoS of NSs efficiently \cite{panotopoulos2022charged}. On the other hand, several cosmological characteristics of the cosmos have been discussed using the generalized polytropic EoS, $p_r = \gamma \rho^{1 + \frac{1}{\eta}} + \beta\rho$ for the first time \cite{chavanis2014models}. Observing that the above-mentioned generalized EoS is unable to characterize the self-bound compact objects, the aforementioned EOS was changed to $p_r = \gamma \rho^{1 + \frac{1}{\eta}} + \beta\rho + \chi$ \cite{azam2015cracking, naeem2021generalized, azam2016study}. To explain the many properties of compact stars, the composition of astronomical compact star models heavily relies on the polytropic EoS. An extensive investigation was carried out by Herrera and Barreto for Newtonian polytropic models in the case of anisotropic fluids \cite{herrera2013newtonian}. The physical effects of the polytropic EoS on charged anisotropic compact star models have also been examined by Takisa and Maharaj \cite{takisa2013some}. The realistic characteristics of uncharged compact star models in the polytropic EoS regime were demonstrated by Thirukkanesh and Ragel \cite{thirukkanesh2012exact, thirukkanesh2013class}. Singh et al. \cite{singh2022anisotropic} recently studied an isotropic solution for polytropic stars in 4$\mathcal{D}$ Einstein-Gauss-Bonnet gravity.
 One can utilize the Einstein field equation to find the solution for a known configuration of matter, but an alternate method of solving the same Einstein equations was explored by assuming the geometry provided by Finch-Skea in the absence of a known matter configuration. Since the matter field and its geometry can have a combined refinement according to the Einstein field equations, we will confirm that a structural strategy, such as the more suitable metric potential will be determined by ensuring an authentic form of that one along with a clear attribution of an analogous metric to address this limitation. Finch and Skea created this technique for the composition of an interior spheroidal geometry \cite{nazar2023relativistic}. \\
 In the present study, we have considered 5$\mathcal{D}$ EGB gravity and a polytropic EoS of type $p_r = \gamma \rho^{1 + \frac{1}{\eta}} + \beta\rho + \chi$ together with the Finch-Skea metric to fully conclude the system of equations. The outline of this manuscript is as follows.
The fundamental field equations and the interior spacetime are discussed in Section \ref{Sec2}. Stellar configurations, i.e., outer spacetime, the equation of state, and system solutions, are covered in the following section. Our inner spacetime has been smoothly matched to the outer Schwarzschild line element to obtain the values of the unknown parameters in this section. A physical investigation of the results is presented in Section \ref{Sec4}, where we also discuss the consistency of the metric coefficients, density, pressure, redshift, and the EoS parameter in the polytropic star model. We examine the stability of our proposed model in Section \ref{Sec5}. Lastly, Section \ref{Sec6} contains closing remarks.

\section{Preliminaries: theoretical setup}\label{Sec2}
Unlike in the Einstein case, a modified action is required in EGB gravity to generate the field equations.
EGB gravity is now expressed in $\mathcal{D}(\geq 5)$ dimensions using the following total action as \cite{maeda2008generalized}:
\begin{eqnarray}\label{action}
I_\mathcal{T} = \frac{1}{2 {\kappa} }\int d^{\mathcal{D}}x\sqrt{-g}\left[ R-2\Lambda +\alpha \mathcal{L}_{\textbf{GB}} \right]+ S_{\mathcal{M}},  
\end{eqnarray}
where $\kappa = 8\pi$, $g$ indicates the determinant of the metric tensor $g_{ij}$, and $R$ denotes the Ricci curvature scalar. The widely accepted cosmological constant is denoted as "$\Lambda$" and the Lagrangian density associated with the GB term is $\mathcal{L}_{\textbf{GB}}$. The action associated with the matter field is represented by $S_{\mathcal{M}}$, and a free coupling constant of dimension $[length]^2$ denoted by $\alpha$ generally known as the Gauss-Bonnet coupling constant. The Gauss-Bonnet coupling constant ($\alpha$) plays a crucial role in Einstein-Gauss-Bonnet gravity, a higher-dimensional extension of GR. It influences the formation of a shock cone and its oscillation properties. This coupling constant mainly includes the quadratic-curvature Gauss-Bonnet term and other higher-dimensional terms in the action. Increasing $\alpha$ causes strong oscillations within the shock cone, potentially leading to Quasi-Periodic Oscillations (QPOs). While negative $\alpha$ values can result in interesting physical outcomes, including changes in the accretion rate and oscillation. For example, increasing $\alpha$ in a negative direction causes the shock opening angle to decrease, whereas increasing $\alpha$ in a positive direction causes it to slightly increase. Also, this Gauss-Bonnet coupling constant ($\alpha$) affects quantum field theory modeling, including modifying the Kovtun-Son-Starinets (KSS) viscosity bound. It also has implications for black hole thermodynamics and dual field theory.
For Minkowski spacetime to be stable in EGB theory, the coupling constant $\alpha$ must be taken into account as positive definite \cite{maeda2007matter}. Furthermore, in string theory, the coupling constant $\alpha$, which is associated with the inverse string tension, is seen as a positive number \cite{boulware1985string}. In our work, we consider  $\alpha > 0$, but a few authors choose to consider both scenarios of $\alpha > 0$ and $\alpha < 0$. Here, we investigated how the Gauss-Bonnet coupling constant, $\alpha$, affected the primary physical properties of our model.\\
In this study, we will now concentrate on five dimensions, i.e., $\mathcal{D}=5$. The gravitational constant, $G$, and the speed of light, $c$, are set to unity by the use of geometric units. 
It should be noted that the auxiliary coupling constant, which represents the UV adjustments to Einstein's theory of GR \cite{Sepul} and is connected to string tension in string theory, develops with the length-square dimension and only takes nonnegative values \cite{Maeda2006final}.
Moreover, a description of the GB term $\mathcal{L}_{\textbf{GB}}$ is given in the expression below: 
\begin{equation}
\mathcal{L}_{\textbf{GB}}=R^{ijkl} R_{ijkl}- 4 R^{ij}R_{ij}+ R^2\label{GB}
\end{equation}
where $R_{ijkl}$ and $R_{ij}$ denote the Riemann curvature tensor and Ricci tensor respectively.
It is important to observe that the Lagrangian with the dimension of length is quadratic in geometric quantities: Ricci tensor, Ricci scalar, and the Riemann tensor.\\
Consequently, the following equations of motion arise from altering the $\mathcal{D}=5$ situation of the previously given action concerning the $g_{ij}$ metric tensor.

\begin{equation}\label{eq3}
G_{ij}+\alpha H_{ij} = \kappa  T_{ij}, 
\end{equation}
where
\begin{eqnarray}
G_{ij} = R_{ij}-\frac{1}{2}R~ g_{ij},
\end{eqnarray}
\begin{eqnarray}
H_{ij} =  2\Big( R R_{ij}-2R_{ik} {R}^k_j -2 R_{ijkl}{R}^{kl} - R_{ikl\delta}{R}^{kl\delta}_j\Big) - \frac{1}{2}~g_{ij}~L_{\textbf{GB}},
\end{eqnarray}
and
\begin{equation}\label{eq3b}
{T}_{ij}=-\frac{2}{\sqrt{-g}}\frac{\delta\left(\sqrt{-g}S_{\mathcal{M}}\right)}{\delta g^{ij}}. 
\end{equation}

$S_{\mathcal{M}}$ produces the energy-momentum tensor $T_{ij}$, which corresponds to the matter field.
In our model, we use the following energy-momentum tensor for the stellar fluid: 
\begin{equation}\label{eq3i}
{T}_{ij}=diag(-\rho^{eff}, p_r^{eff}, p_t^{eff}, p_t^{eff}, p_t^{eff}),
\end{equation}
where $\rho^{eff} = \rho$, $p_r^{eff} = p_r$ and $p_t^{eff} = p_t$.\\
$\rho$, $p_r$, and $p_t$ indicate the matter-energy density, radial pressure, and the tangential pressure.\\
At this point, we can explore adding a static, spherically symmetrical $5\mathcal{D}$ geometry to the EGB framework's governing equations.
We assume that the interior of our stellar object is illustrated by the following line element in coordinates $(\xi = t, r, \theta, \phi, \psi)$:
 \begin{eqnarray}
\label{5} ds^{2}_{\bold{int}}= -e^{2\nu} dt^{2} + e^{2\lambda} dr^{2} +
 r^{2}(d\theta^{2} + \sin^{2}{\theta} d\phi^2 +\sin^{2}{\theta} \sin^{2}{\phi}
 d\psi^2),
\end{eqnarray}
where $\nu$ and $\lambda$ are solely function of radial coordinate $r$ only.
 \\
Thus, the remaining components of the metric tensor $g_{ij}$ and its inverse $g^{ij}$ are as follows:
\begin{eqnarray}\label{5b} 
g_{ij}=-e^{2\nu}\delta_i^0\delta_j^0+e^{2\lambda}\delta_i^1\delta_j^1+r^2\delta_i^2\delta_j^2+r^2\sin^{2}{\theta}(\delta_i^3\delta_j^3+\sin^{2}{\phi}\delta_i^5\delta_j^5),
\end{eqnarray}
\begin{eqnarray}\label{5c} 
g^{ij}=e^{-2\nu}\delta^i_0\delta^j_0+e^{-2\lambda}\delta^i_1\delta^j_1+\frac{1}{r^2}\delta^i_2\delta^j_2+\frac{1}{r^2 \sin^{2}{\theta}}\left(\delta^i_3\delta^j_3+\frac{1}{\sin^{2}{\phi}}\delta^i_5\delta^j_5\right).
\end{eqnarray}
Assuming that the five-velocity in the EGB system is given by $u^i=e^{-\nu}\delta_0^i$ and using the preceding metric, we may construct the following set of independent equations,
\begin{eqnarray}
\label{7a}\kappa\rho^{eff}= \kappa \rho  &=& -\frac{3}{r^3 e^{4\lambda }} \Bigg[-re^{4\lambda} + re^{2\lambda} -4\alpha e^{2\lambda}\lambda' - r^2 e^{2\lambda}\lambda'  + 4\alpha \lambda'\Bigg],\label{fe1}\\
\label{7b} \kappa p_r^{eff} = \kappa p_r & = &\frac{3}{r^3 e^{4\lambda }}
\Bigg[-re^{4\lambda} + (r^2 \nu' +r +4\alpha \nu')e^{2\lambda} -4\alpha \nu'\Bigg] , \label{fe2}\\
\label{7c} \kappa p_t^{eff} = \kappa p_t &=& 
 \frac{1}{r^2 e^{2\lambda }} \Bigg[r^2(\nu')^2 + 2r \nu'
-2r\lambda' - r^2\nu'\lambda'  + 1 \Bigg] +\frac{1}{r^2 e^{2\lambda}} \Bigg[ 4\alpha (\nu')^2  +4\alpha \nu'' + r^2 \nu'' -4\alpha\nu'\lambda' \Bigg] \nonumber \\ && + \frac{1}{r^2 e^{4\lambda }} \Bigg[-e^{4\lambda} + 12 \alpha \nu' \lambda' -4 \alpha(\nu')^2 - 4\alpha \nu''\Bigg].\label{fe3}
\end{eqnarray}
Here the prime symbol $(')$ represents the differentiation with regard to the radial coordinate `$r$'.

\section{Stellar configuration}\label{Sec3}

Basic junction conditions guarantee smooth matching at the interface between the exterior and interior spacetimes of a stationary celestial object. Moreover, even though the surface of a celestial structure should not have any radial pressure, the boundary does not always release tangential pressure. These factors compel an extensive investigation of the junction conditions of a neutron star. There are numerous methods available in the literature for characterizing the inner region of a compact star structure. We have taken into account the generalized Finch-Skea {\em ansatz} \cite{finch1989realistic, maurya2011family, patel2023generalized} in order to build a stellar model. The mathematical form of the {\em ansatz} is given by,
\begin{equation}\label{lam}
2\lambda = ln(1 + Ar^2)^n,    
\end{equation}

where `$A$' represents arbitrary constant parameters with the unit in $km^{-2}$.  Moreover, $n( > 0)$ is a dimensionless parameter. For simplicity, here we consider $n=2$. \\
To obtain the constant parameter $A$ for our proposed model, a suitable static and spherically symmetric exterior vacuum Schwarzschild formulation must be matched with the interior space-time solution. Furthermore, the Schwarzschild vacuum solution is essential to astrophysics due to its asymptotically flat aspect.
In the $5\mathcal{D}$ EGB formalism, the most acceptable exterior solution is provided by Glavan and Linis \cite{glavan2020einstein} and is given by: 

\begin{eqnarray}\label{ext}
ds^{2}_{\bold{ext}} &=& -\mathfrak{f}(r)dt^2 + \frac{dr^2}{\mathfrak{f}(r)} + r^2(d\theta^2+\sin^2\theta d\phi^2 + \sin^2\theta\sin^2\phi^2 d\psi^2), \label{metric}
\end{eqnarray}
where 
\begin{eqnarray}
\mathfrak{f}(r) = 1 + \frac{r^2}{32\pi\alpha}\left(1 \pm \sqrt{1 + \frac{128\pi\alpha \mathcal{M}}{r^3}}\right), 
\end{eqnarray}
Here,  `$\mathcal{M}$' indicates the gravitational mass of the star. It can be shown that the negative branch is preferred in five-dimensional spacetime \cite{singh2022anisotropic}. So, we consider only $$\mathfrak{f}(r) = 1 + \frac{r^2}{32\pi\alpha}\left(1 - \sqrt{1 + \frac{128\pi\alpha \mathcal{M}}{r^3}}\right).$$
We claim that, for a compact polytropic star, the EGB-Sch solution and the star's intrinsic solution relate exactly in this case, enabling us to compare the interior metric with the EGB-Sch exterior vacuum spacetime.

\subsection{Assumption of EoS}\label{eosq}
Despite the logical clarity of the system of equations, obtaining explicit solutions to the field equations described above (\ref{f1})-(\ref{f3}) is too challenging because there are only three equations and five unknowns, $\{\rho, ~p_r, ~p_t, ~\nu, ~\lambda\}$.
To avoid this complication, we have implemented one assumption: we have assumed a particular form of $\lambda$. We utilized the polytropic EoS provided by
\begin{equation}\label{eos1}
p_r = \gamma \rho^{1 + \frac{1}{\eta}} + \beta\rho + \chi,    
\end{equation}
i.e., a convincing nonlinear relationship arises between the density of normal baryonic matter $(\rho)$ and the radial pressure $p_r$. Here $\eta$ is the polytropic index, and $\beta$, $\gamma$, and $\chi$ are constant parameters with proper dimensions. The polytropic index has been set to one, or $\eta = 1$, to obtain an exact solution; i.e. the EoS becomes 
\begin{equation}\label{eos2}
p_r = \gamma \rho^2 + \beta\rho + \chi.    
\end{equation}
The EoS (\ref{eos2}) features a quadratic contribution $\gamma\rho^2$, which typically expresses the neutron liquid in Bose-Einstein condensate form, and the linear term $\beta\rho + \chi$ originates from the free quarks model of the well-known MIT bag model, with specific values of $\beta = 1/3$ and $\chi = -4B_g/3$, where $B_g$ is the bag constant. 

\subsection{Proposed stellar model}

The numerical value of the constant will be derived by a smooth matching of our chosen interior and exterior solutions.
Using the expressions given in (\ref{eos1}),(\ref{fe1}) and (\ref{fe2}), we obtain a non-linear differential equation upon simplification as follows: 
\begin{eqnarray}\label{dnu}
 \nu'(r) &=& -r (1 + A r^2)^4 \Bigg[-\frac{3}{r^2} - \frac{3 \beta}{r^2} + \frac{24 A \alpha \beta}{r^2 (1 + A r^2)^5} - \frac{6 A \beta}{(1 + A r^2)^3} - \frac{24 A \alpha \beta}{r^2 (1 + A r^2)^3} \nonumber\\&& + \frac{3}{(r + A r^3)^2} + \frac{3\beta}{(r + A r^3)^2} - \frac{\Big(9 A^2 \gamma \Big[4 + A \Big\{11 r^2 + 8 \alpha (2 + A r^2) + 
             A r^4 (11 + A r^2 (5 + A r^2))\Big\}\Big]\Big)^2}{8 \pi (1 + A r^2)^{10}}  \nonumber\\&& - 8 \pi\chi\Bigg] \Bigg/ \Bigg[3 \{1 + A (4 \alpha + r^2) (2 + A r^2)\}\Bigg]. 
\end{eqnarray}
 Now integrating equation (\ref{dnu}) we obtain the analytic expression for $\nu(r)$ as,
 \begin{eqnarray}\label{nu}
  \nu(r) &=& \frac{4\pi r^2 \chi}{3} + 
 A \Bigg[(1 + \beta) r^2 - \frac{\gamma \{11 + 3 A r^2 (5 + 2 A r^2)\}}{
    8 \pi (1 + A r^2)^3} + \frac{4 \pi r^2 (-8 \alpha + r^2) \chi}{3}\Bigg] +\nonumber\\&& 
 A^2\Bigg[\frac{3\gamma r^2}{16\pi} + \frac{r^4}{4} + \frac{\beta r^4}{4} - \frac{
    3 \alpha \gamma (9 + 5 A r^2)}{20 \pi (1 + A r^2)^5} + 
    \frac{64 \alpha^2 \pi r^2 \chi}{3} + \frac{4\pi r^6 \chi}{9} \nonumber\\&& 
    -\frac{2 \alpha r^2 (3 + 3 \beta + 4 \pi r^2 \chi)}{3}\Bigg] + 
 \beta ~\ln(1 + A r^2) + \Bigg[\sqrt{A} \{9 \gamma + 
      32 \alpha \pi (-3 - 3 \beta + 32 \alpha \pi \chi)\} \{(1 - 
         8 A \alpha + 8 A^2 \alpha^2 \nonumber\\&& + 
         4 \sqrt{A} \sqrt{\alpha} \sqrt{-1 + A \alpha} - 
         8 A^{3/2} \alpha^{3/2} \sqrt{-1 + A \alpha}) ~\ln\{
        \sqrt{A} - 2 A \sqrt{\alpha} \sqrt{-1 + A \alpha} + 
         A^{3/2} (2 \alpha + r^2)\} \nonumber\\&& + (-1 + 8 A \alpha - 8 A^2 \alpha^2 + 
         4 \sqrt{A} \sqrt{\alpha} \sqrt{-1 + A \alpha} - 
         8 A^{3/2} \alpha^{3/2} \sqrt{-1 + A \alpha})~ \ln(
        \sqrt{A} + 2 A \sqrt{\alpha} \sqrt{-1 + A \alpha} \nonumber\\&& + 
         A^{3/2} (2 \alpha + r^2))\}\Big/(192\pi \sqrt{\alpha}
     \sqrt{-1 + A \alpha})\Bigg] + C,  
 \end{eqnarray}
where `$C$' is the integration constant.

So, from the field equations (\ref{fe1})-(\ref{fe3}) $\rho$, $p_r$ and $p_t$ can be taken in the following forms for the interior part of the star:
\begin{eqnarray}
\rho  &=& \frac{3 A [4 + A \{11 r^2 + 8 \alpha (2 + A r^2) + 
      A r^4 (11 + A r^2 (5 + A r^2))\}]}{8 \pi (1 + A r^2)^5}, \label{f1} \\
p_r  &=& \frac{3 A \beta [4 + A \{11 r^2 + 8 \alpha (2 + A r^2) + 
       A r^4 (11 + A r^2 (5 + A r^2))\}]}{8 \pi(1 + A r^2)^5}\nonumber\\&&  + \frac{9 A^2 \gamma [4 + 
    A\{(11 r^2 + 8 \alpha (2 + A r^2) + 
       A r^4 (11 + A r^2 (5 + A r^2))\}]^2}{64 \pi^2 (1 + A r^2)^{10}} + \chi , \label{f2}\\
p_t  &=& \frac{f_1(r) +f_2(r)}{[4608 \pi^3 (1 + A r^2)^{16} \{1 + A (4 \alpha + r^2) (2 + A r^2)\}]},\label{f3}
 \end{eqnarray}
where the functions $f_1$ and $f_2$ are given in the 'Appendix' section.\\
At the boundary $r = \mathfrak{R}$, when we match the interior and exterior spacetimes, we obtain
$$e^{2\nu(\mathfrak{R})} = e^{-2\lambda(\mathfrak{R})} = \mathfrak{f}(\mathfrak{R}).$$
So, $e^{-2\lambda(\mathfrak{R})} = \mathfrak{f}(\mathfrak{R})$ gives
\begin{eqnarray}
A &=& \frac{\Big[\frac{1}{\sqrt{\mathfrak{f}(\mathfrak{R})}} - 1\Big]}{\mathfrak{R}^2} \nonumber\\
&=& \frac{-1 + \frac{1}{\sqrt{
 1 + \frac{[1 - \sqrt{1 + (128 \alpha \mathcal{M} \pi)/\mathfrak{R}^3}] \mathfrak{R}^2}{
  32 \alpha \pi}}}}{\mathfrak{R}^2}.
\end{eqnarray}
Equation (\ref{nu}) and $e^{2\nu(\mathfrak{R})} = \mathfrak{f}(\mathfrak{R})$ give the value of $C$ such that
\begin{eqnarray}
 C &=& \frac{1}{2}ln[\mathfrak{f}(\mathfrak{R})] - \nu(\mathfrak{R})   
\end{eqnarray}
 Additionally, the radial pressure must disappear at the boundary i.e. $p_r(\mathfrak{R}) = 0$, which results
\begin{eqnarray}
\chi &=& -\gamma\{\rho(\mathfrak{R})\}^2 - \beta\rho(\mathfrak{R}) \nonumber\\
&=& \frac{1}{64 \pi^2 (1 + A \mathfrak{R}^2)^{10}} \Big[3 A [4 + 
   A (11 \mathfrak{R}^2 + 8 \alpha (2 + A \mathfrak{R}^2) + 
      A \mathfrak{R}^4 \{11 + A \mathfrak{R}^2 (5 + A \mathfrak{R}^2))\}] [-8 \beta \pi (1 + A \mathfrak{R}^2)^5 \nonumber\\&& -3 A \gamma \{4 + A (11 \mathfrak{R}^2 + 8 \alpha (2 + A \mathfrak{R}^2) + 
         A R^4 (11 + A \mathfrak{R}^2 (5 + A \mathfrak{R}^2)))\}]\Big].
\end{eqnarray}
\section{Celestial features}\label{Sec4}

To obtain more insight into the physical characteristics of the generated celestial object, we need to identify certain suitable values for the supporting model parameters.
We consider the compact star candidate EXO 1785-248 in this work to determine appropriate values for the free parameters $A$ and $\chi$.
The mass and radius values of the strange spherical object EXO 1785-248 are $\mathcal{M} = 1.3 \pm 0.2~\mathcal{M}_{\odot}$ and $\mathfrak{R} = 8.849_{-0.4}^{+0.4}$ km \cite{gangopadhyay2013strange}. We set the model parameters $\gamma$ and $\beta$ to $\gamma = 16.9$ and $\beta = 0.31$, respectively, to build a physically valid and realistic stellar model. Finally, utilizing the recently obtained data of the EXO 1785-248 in addition to various coupling parameters $\alpha \in [4, 8]$, we derive the numerical findings shown in Table \ref{table1}.

\begin{table}[h]
\centering
\caption{\label{table1} Several plausible values of the free model parameters}
\begin{tabular}{|c|c|c|}
\hline
$\alpha$ & $A~(\text{km}^{-2})$  & $\chi$ \\
\hline
\hline
 4 & 0.00209931 & -0.000237665 \\
 5 & 0.00193297 & -0.000224617 \\
 6 & 0.00180199 & -0.000213899 \\
 7 & 0.00169525 & -0.000204864 \\
 8 & 0.00160602 & -0.000197097 \\
\hline
\end{tabular}
\end{table}

\subsection{Regularity of the metric potentials}\label{mp}
Metric potentials inside the star must not contain singularities to find an acceptable and realistic model. The metric potential values at the center of the stellar object can easily be verified as follows:\\
$\lim_{r\rightarrow 0}e^{\nu}= $ a non-zero constant, and $\qquad \lim_{r\rightarrow 0}e^{\lambda}=1.$  \\ 
As a consequence of these findings, the metric potentials in the fluid configuration are singularity-free at the center of the structure and have finite values.
Furthermore, at the center of the star, we have,
$\frac{de^{\nu}}{dr}\Big\rvert_{r=0}=0$
and
$\frac{d e^{\lambda}}{dr}\Big\rvert_{r=0} = 0$. Furthermore, within the star, the derivatives of the metric potential components are positive and consistent. Moreover, $\frac{d^2 e^{\nu}}{d^2 r} \Big\rvert_{r=0} >0,$ and  
$\frac{d^2 e^{\lambda}}{d^2 r} \Big\rvert_{r=0} >0.$\\
Therefore, $e^{\nu}$ and $e^{\lambda}$ assume minimum values close to the core and progressively increase with $r$. In the range $(0, \mathfrak{R})$, we can verify that the metric coefficients behave as expected because the metric components under discussion are non-singular at their center.
Considering the radial profiles of the metric coefficients, which are shown in Fig.~\ref{metric}, this can be easily verified.
\begin{figure}[H]
    \centering
        \includegraphics[scale=.63]{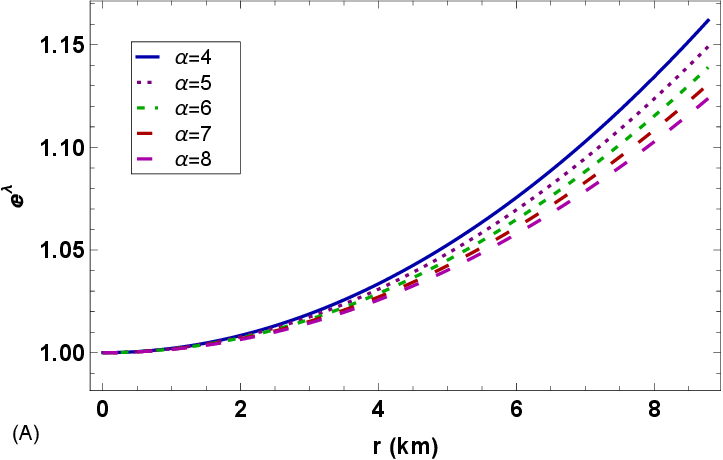}
         \includegraphics[scale=.63]{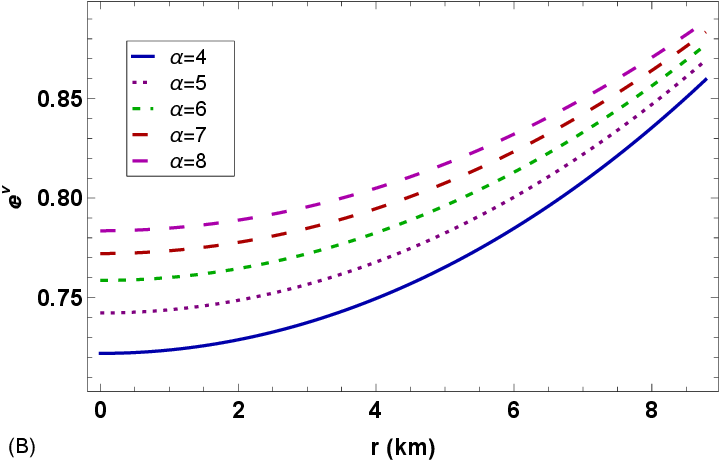}
        \caption{Variation of metric functions $e^{\lambda(r)}$ and $e^{\nu(r)}$ with respect to $r$.}\label{metric}
\end{figure}

\subsection{Regularity of the fluid components}
Physical parameters such as the energy density $\rho$ and anisotropic stresses, i.e., $p_r$ and $p_t$, have a major impact on the evolutionary change of very dense strange star configurations. The aforementioned influencing factors must have finite and non-singular values to show their viability at the center. In addition, the radial pressure of the star should also disappear at its boundary, or $p_r (r = \mathfrak{R}) = 0$.
In this regard, the pressure and core energy density are determined as follows:
\begin{equation}\label{rhoc} 
\rho\Big\rvert_{r=0} = \rho_c = \frac{3 A (4 + 16 A \alpha)}{8\pi},
\end{equation}
\begin{equation}\label{pc} 
 p_c = p_r\Big\rvert_{r=0} =\frac{(9 A^2 (4 + 16 A \alpha)^2 \gamma)}{64 \pi^2} + \frac{3 A (4 + 16 A \alpha) \beta}{8 \pi} + \chi.
\end{equation}
In furtherance of the previously mentioned results, the surface energy density can be computed using the following relation:
\begin{eqnarray}\label{rhos} 
\rho\Big\rvert_{r=\mathfrak{R}} =\rho_s = \frac{3A[4 + A \{11 \mathfrak{R}^2 + 8 \alpha (2 + A \mathfrak{R}^2) + 
      A \mathfrak{R}^4 (11 + A \mathfrak{R}^2 (5 + A \mathfrak{R}^2))\}]}{8\pi (1 + A \mathfrak{R}^2)^5}.   \label{ro}
\end{eqnarray}

\begin{figure}[H]
    \centering
        \includegraphics[scale=.63]{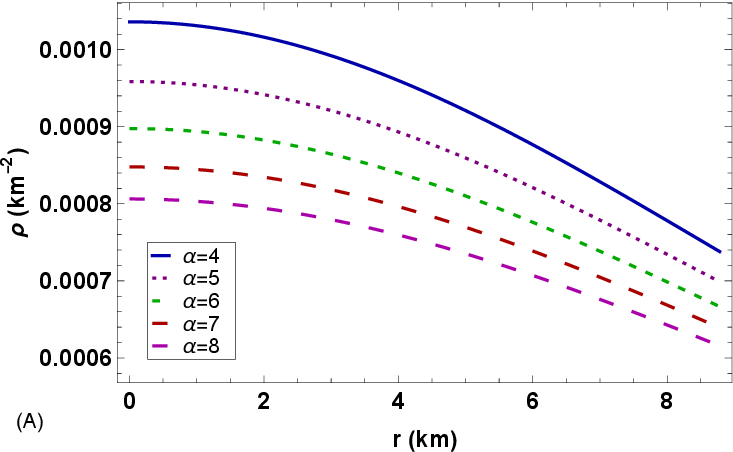}
         \includegraphics[scale=.63]{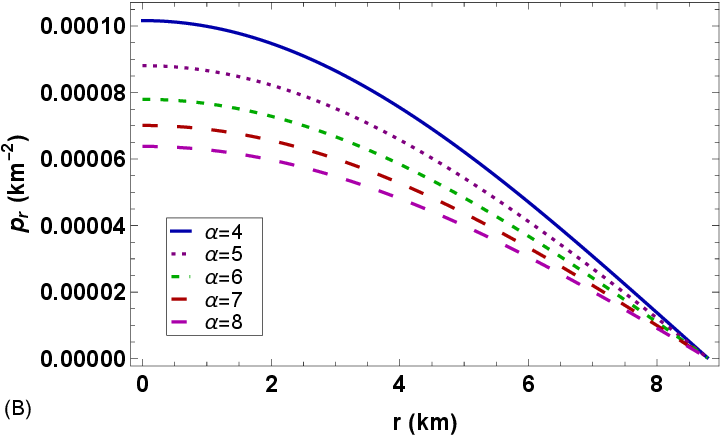}
         \includegraphics[scale=.63]{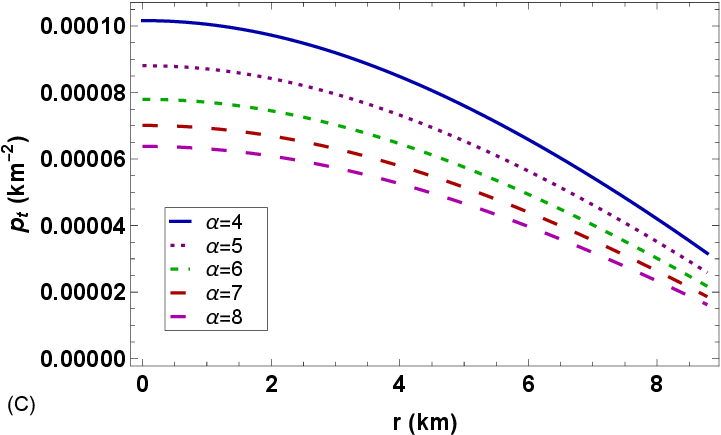}       
        \caption{Profiles of (A) matter density and (B, C) pressure components with respect to $r$.}\label{rho}
\end{figure}

The density and pressure profiles for different values of $\alpha$ are shown in Fig.~\ref{rho}, and the analysis suggests that all of them are monotonically decreasing functions of $r$'. This means that at the boundary, where the radial pressure $p_r$ vanishes, and at the center of the star, where they are at their highest value.
At both the stellar surface, $r = \mathfrak{R}$, and the center, $r = 0$, the tangential pressure $p_t$, has non-zero finite values and decreases monotonically with $r$.

In Table \ref{table2}, we give some pertinent numerical values for the central energy density $\rho_c$ and the central radial pressure $p_c$.

\begin{table}[h]
\centering
\caption{\label{table2} Numerical measurements of the surface energy density and the central fluid components.}
\begin{tabular}{|c|c|c|c|c|}
\hline
$\alpha$ & $\rho_c \times 10^{14}$ ($\text{gm}~\text{cm}^{-3}$) & $\rho_s\times 10^{14}$ ($\text{gm}~\text{cm}^{-3}$) & $p_c\times 10^{14}$ ($\text{gm}~\text{cm}^{-3}$) \\
\hline
\hline
 4   & 1.39792 & 9.94516   & 1.37143 \\
 5   & 1.29347 & 9.41842   & 1.18850 \\
 6   & 1.21115 & 8.98419   & 1.05209 \\
 7   & 1.14402 & 8.61705   & 0.94608 \\
 8   & 1.08786 & 8.30061   & 0.86112 \\
\hline
\end{tabular}
\end{table}

\subsection{Nature of the fluid components}
\begin{figure}[H]
    \centering
        \includegraphics[scale=.63]{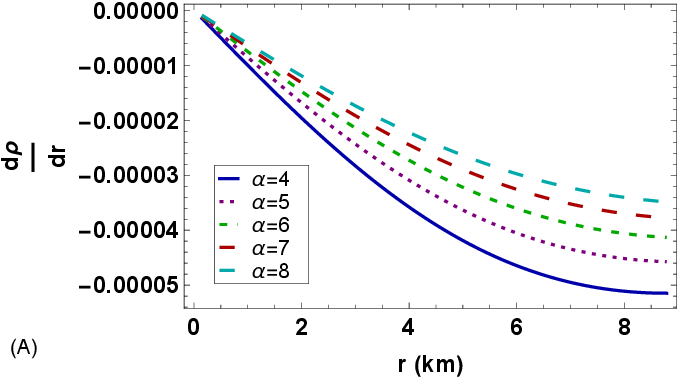}
         \includegraphics[scale=.63]{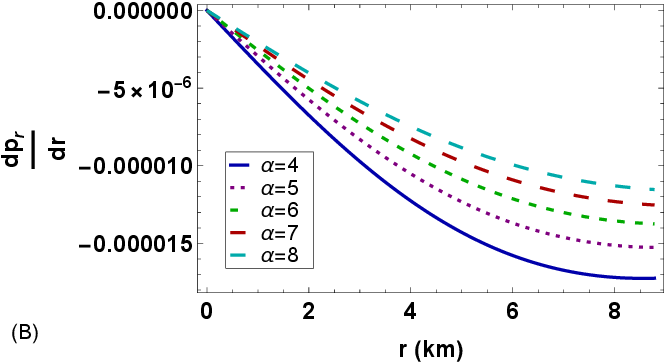}
         \includegraphics[scale=.63]{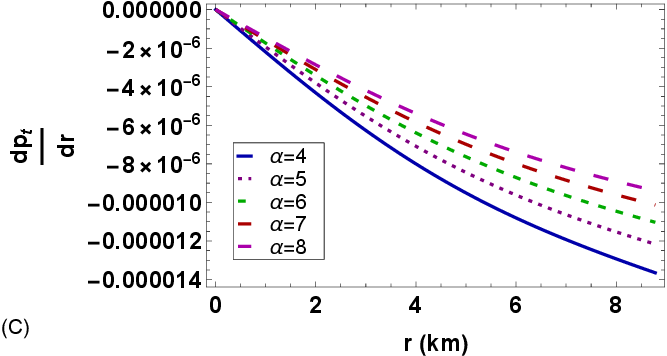}
         \caption{Gradients of the matter density and pressure components with respect to $r$.}\label{grad}
\end{figure}

To test the validity of our model, which describes an anisotropic compact celestial structure, we proceed with the analysis of a few specific aspects of the celestial layouts, namely $\frac{d\rho}{dr}$, $\frac{dp_r}{dr},$ and $\frac{dp_t}{dr}$, utilizing a graphic analysis of density and pressure gradients. Observing the pattern of change in the pressure and energy gradients in Fig.~\ref{grad}, we conclude that the presence of an intense celestial structure is ensured by their persistent negative behaviors inside the star.
We also point out that the energy density and pressure gradients decrease with increasing parameter $\alpha$. Thus, the discussed stellar model also meets the requirements for a realistic star configuration, namely $\frac{d\rho}{dr} < 0, \frac{dp_r}{dr} < 0,$ and $\frac{dp_t}{dr} < 0$.

\begin{figure}[H]
    \centering
         \includegraphics[scale=.63]{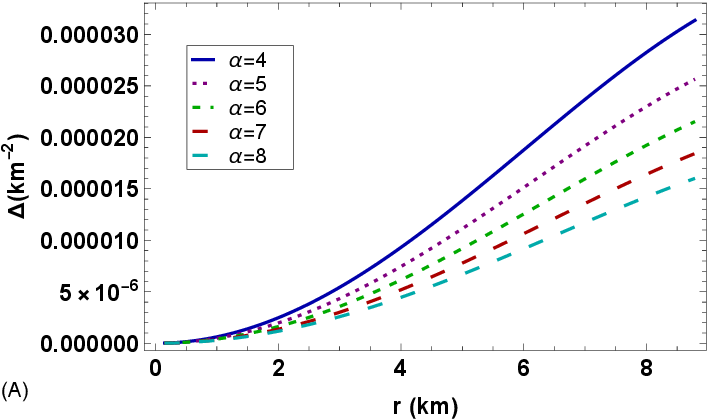}
         \includegraphics[scale=.63]{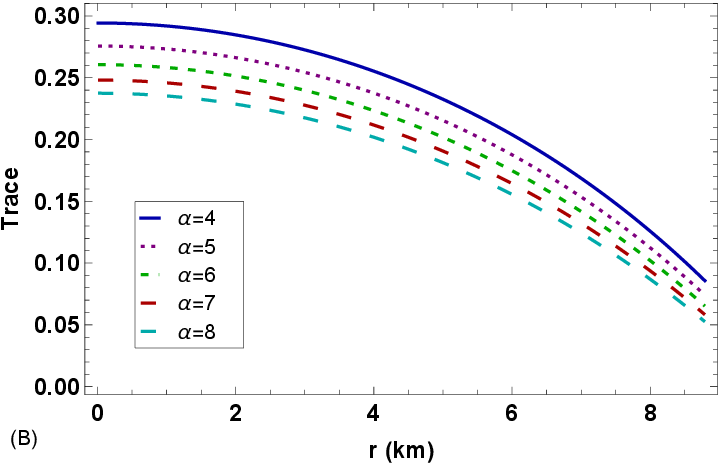}
        \caption{(A) Anisotropic factor ($\Delta$) and (B) the trace profile with respect to $r$.}\label{delta}
\end{figure}

We now examine the implications of pressure anisotropy, or $\Delta = p_t - p_r$, on the compact celestial object. The transverse pressure ($p_t$) becomes non-zero near the star's boundary, while the radial pressure ($p_r$) disappears at that point.
$p_r$ and $p_t$ are identified as equal at the center, but $p_t > p_r$ has been found away from the center, which causes anisotropy.
Figure~\ref{delta} makes it evident that the positive anisotropy, or $p_t > p_r$, relates to the anisotropic force gradient in the star's repulsive nature, which plays a role in counterbalancing the gravitational force gradient and creating equilibrium and stability in the model. In our current model, the type of pressure anisotropy for $4 \leq \alpha \leq 8$ is shown in the left panel of Fig. \ref{delta}. The anisotropic factor disappears at the star's center ($r = 0$) and increases with distance from the boundary ($r = \mathfrak{R})$ as $\Delta_{(r>0)} > 0 $. The direction of the pressure anisotropy, which depends on the pressure components, becomes an interesting and important feature when we examine the physical structure. When $\Delta < 0$, that is, when $ p_t < p_r $, the anisotropy is attractive or directed inward. On the other hand, it is assumed that the anisotropy is repulsive or directed outward when $\Delta > 0$, that is, when $p_t > p_r $\cite{hossein2012anisotropic1}. 
Since the radial and transverse pressures are identical at the stellar core, the pressure anisotropy should vanish there, suggesting that the pressure has become isotropic.
Positive anisotropy within the stellar structure is also required because it generates a repulsive force that keeps the star from collapsing under the force of gravity \cite{gokhroo1994anisotropic}. For a physically acceptable model, $\Delta_{(r=\mathfrak{R})} > 0$ is always positive since $p_t(r=\mathfrak{R}) > 0 $ and $p_r(r=\mathfrak{R}) = 0$ and it achieves its highest position from the center towards the boundary surface.\\
On the right panel of Fig. \ref{delta}, we display the trace profile, or $(\frac{p_r+2p_t}{\rho})$, of the celestial structure. More importantly, it is positive across the fluid sphere and declines monotonically with increasing radial coordinates, with the highest values close to the stellar core. The result corresponds with a considerable profile of stellar matter variables, which could be interpreted as the compact environment surrounding a possible celestial configuration \cite{rej2023charged}.

\subsection{Mass-radius ratio and compactness}
\begin{figure}[H]
    \centering
\includegraphics[scale=.63]{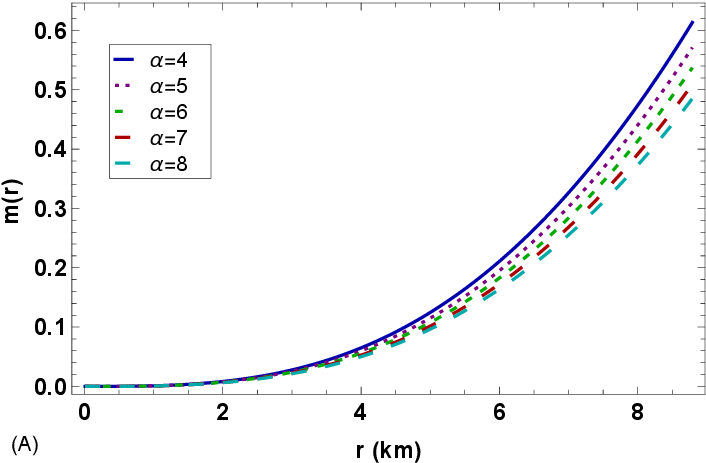}
\includegraphics[scale=.63]{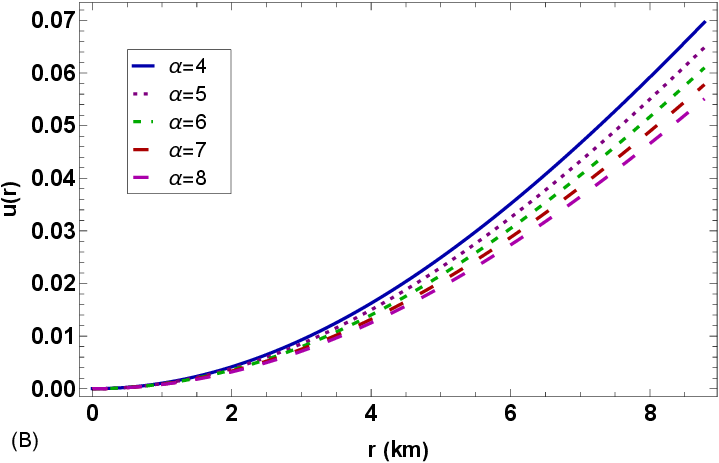}
\caption{(A) Mass function and (B) compactness factor against $r$.}\label{massu}
\end{figure}

Buchdahl \cite{Buchdahl1959general} suggested that there is an upper limit for the ratio of acceptable maximum mass to radius $\frac{2\mathcal{M}}{\mathfrak{R}} < \frac{8}{9}$ for a spherically symmetric static strange star composed of a perfect anisotropic fluid. $\frac{2\mathcal{M}}{\mathfrak{R}} = 0.4334 < \frac{8}{9}$ in this scenario, again resulting in a stable star model.
It is possible to determine the mass function $m(r)$ by calculating the following integral in the given system, as follows:

\begin{eqnarray}\label{mm}
 m(r)=4\pi\int^r_0{\rho \xi^2 d\xi} 
 \end{eqnarray}
Using the above expression, we can calculate the gravitational mass $\mathcal{M}$ at the boundary of our star as follows:
\begin{eqnarray}\label{mm1} 
\mathcal{M}= m(r)\Big\rvert_{r=\mathfrak{R}}=\frac{\mathfrak{R}}{2}\Big[1-\frac{1}{(1 + A\mathfrak{R}^2)^2}\Big].
\end{eqnarray}
Additionally, the effective mass can be computed as
\begin{eqnarray}\label{mef}
\mathcal{M}^{eff} = 4\pi\int^{\mathfrak{R}}_0{ \rho \xi^2 d\xi} .
\end{eqnarray}
The compactness factor $u(r)$ and its actual form are defined as follows:

\begin{eqnarray}\label{ur}
 u(r) = \frac{m(r)}{r},   
\end{eqnarray}
and
\begin{eqnarray}\label{ur2}
 u^{eff}=\frac{\mathcal{M}^{eff}}{\mathfrak{R}}.    
\end{eqnarray}
It should be mentioned that for white dwarfs, $u = 10^{-3}$, and for ordinary stars, $u = 10^{-5}$. Further predictions state that $u = 0.5$ for a black hole, $u \in (0.1, 0.25)$ for a strange neutron star, and $u \in (0.25, 0.5)$ for an ultra-compact stellar object. Fig. \ref{massu} illustrates the behaviors of the compactness factor and the mass function. It is evident that each of the aforementioned numbers is positive and regular, and increases monotonically as $r$ increases within the celestial material. Furthermore, as the coupling parameter $\alpha$ rises, $u(r)$ falls.

\subsection{Redshift functions}
Redshift is a very important tool in cosmology and astronomy because it makes it easy to investigate the properties of our galaxy and ultimately the entire cosmos. More appropriately, the term `cosmological redshift' refers to what was formerly known as the Doppler shift in the context of an expanding universe. Since the redshift, or fractional change in the wavelength of emitted and received light, can be measured, cosmologists typically use it to refer to past events rather than the period at which they occurred. One may only refer to an estimated time by assuming that the cosmos evolved throughout time, which is not measurable. The most effective approach is utilizing an expansion model concerning a specific scenario, such as the Friedmann and Lemaitre equations, with specific assumptions made on the cosmological parameters that are included in it. There are three known parameters, the curvature parameter of the universe $R$, Einstein's cosmological constant $\Lambda$, and the densities of matter and radiation $\rho$, compared to one known parameter, the redshift $Z$. These parameters are only known within errors. The fractional difference in the wavelength of light between that received by the observer $\lambda_0$ and that released by source $\lambda$ is the redshift of light emitted by a receding source, for instance, a galaxy or a spherically symmetric static compact object, i.e.\cite{glendenning2010special}
\begin{eqnarray}
Z = \frac{\Delta \lambda}{\lambda} = \frac{\lambda_0- \lambda}{\lambda} \label{basic z} 
\end{eqnarray}
So, in the case of a Schwarzschild star, we have 
\begin{eqnarray}
Z = \Bigg(1 - \frac{2\mathcal{M}}{\mathfrak{R}}\Bigg)^\frac{1}{2}- 1 \leq \frac{\mathcal{M}}{\mathfrak{R}} 
\end{eqnarray}
The loss of energy experienced by electromagnetic waves or photons leaving a gravitational field, particularly at the surface of a massive star, causes the gravitational redshift, also known as inner redshift, to shift the electromagnetic radiation of an object towards the less energetic (higher wavelength) end of the spectrum. In contrast, a blueshift, also known as a negative redshift, is characterized by a decrease in wavelength and an increase in frequency and energy. When a photon leaves the center and travels to the surface, it has to pass through the dense core region much more often, which causes additional dispersion and energy loss. However, a photon emitting from close to the surface will have a shorter route via a denser area, leading to reduced dispersion and energy loss. Consequently, the inner redshifts of the surface and center are the lowest and highest, respectively \cite{karmakar2024celestial}. An intriguing feature is that when mass grows, the radius will likewise somewhat increase, leading to a rise in surface gravity and surface redshift. As a result, the trends for the surface and inner redshifts are at odds \cite{rej2023charged}. 

\begin{figure}[H]
    \centering
\includegraphics[scale=.63]{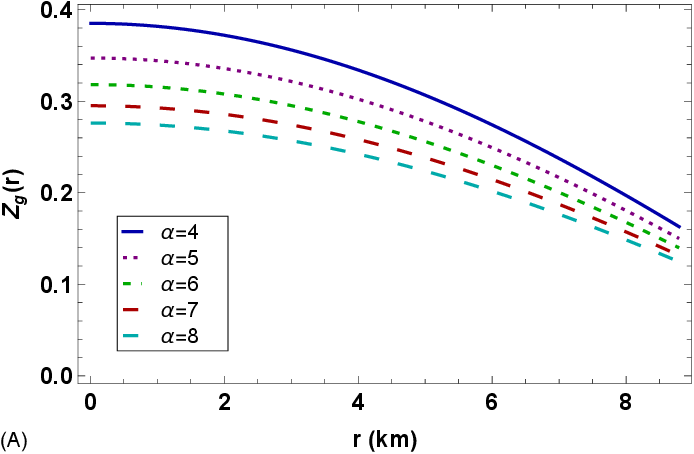}
\includegraphics[scale=.63]{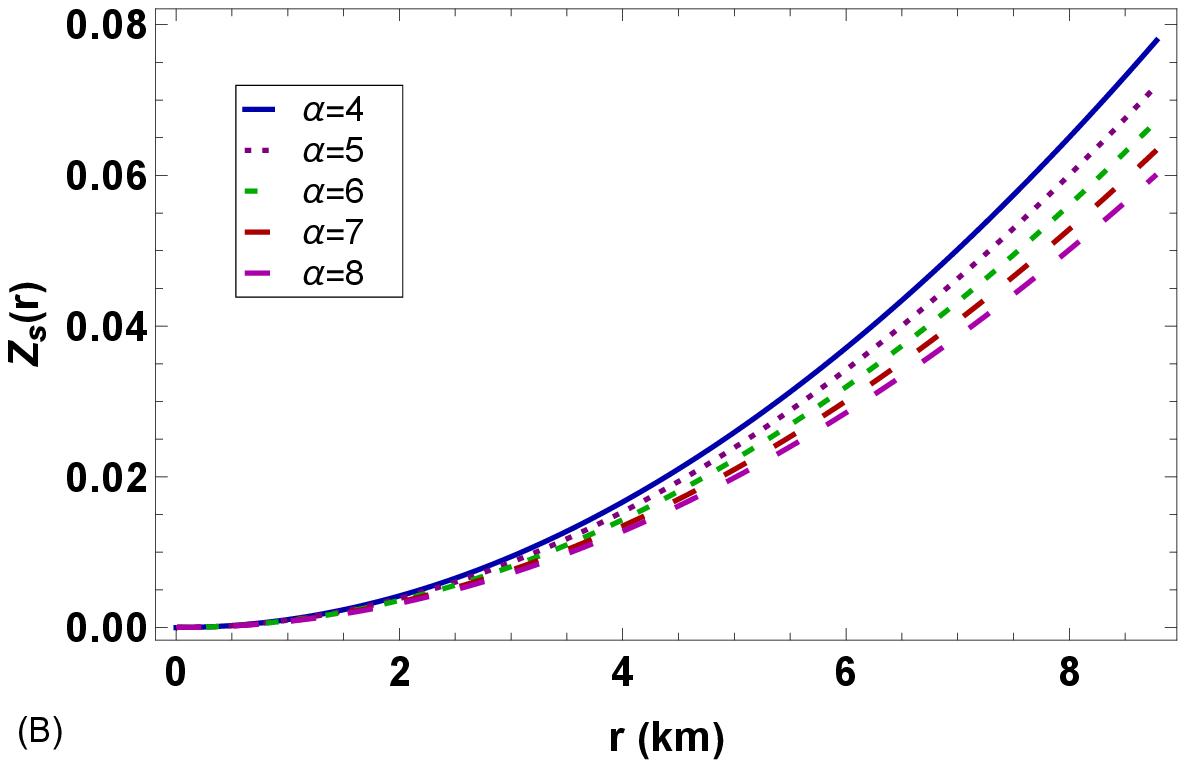}
\caption{Behavior of the (A) inner and (B) surface redshift functions against $r$.}\label{redshift}
\end{figure}

The following expression defines inner redshift $Z_g$ as:
\begin{eqnarray}
   Z_g(r) = e^{-\nu(r)} - 1 
\end{eqnarray}
and equation (\ref{ur}) can be utilized to represent the surface redshift function $Z_s$ as:
\begin{eqnarray}
    Z_s = \frac{1}{\sqrt{1 - 2u(r)}} - 1.
\end{eqnarray}
The maximum surface redshift of an anisotropic fluid sphere is expected to occur\cite{bohmer2006bounds}, with a limitation $Z_s \leq 5.211$. Fig.~\ref{redshift} displays the profiles for both redshifts. It is noticeable that they are regularly positive, with the inner redshift decreasing with $r$' and the surface redshift growing monotonically. 
The surface redshift is valid for our stellar object and satisfies the conditions throughout the stellar configuration, as seen in the right panel of Fig. \ref{redshift}.

In Table~\ref{table3}, for different values of the coupling parameter $\alpha$, we give a comparative numerical analysis of the effective compactness ($u^{eff}(\mathfrak{R})$) and surface redshift function($Z_s(\mathfrak{R})$).

\begin{table}[h]
\begin{center}
\caption{\label{table3} Comparative study of effective compactness and surface redshift.}
\begin{tabular}{|c|c|c|c|}
\hline
$\alpha$   &   $u^{eff}(\mathfrak{R})$  &  $Z_s(\mathfrak{R})$ \\
\hline\hline
4  & 0.615283 & 0.0782256 \\
5  & 0.572880 & 0.0722358 \\
6  & 0.538813 & 0.0674952 \\
7  & 0.510601 & 0.0636166 \\
8  & 0.486699 & 0.0603633 \\
\hline
\end{tabular}
\end{center}
\end{table}

\subsection{EoS parameter: Zeldovich's condition}
\begin{figure}[H]
    \centering
\includegraphics[scale=.63]{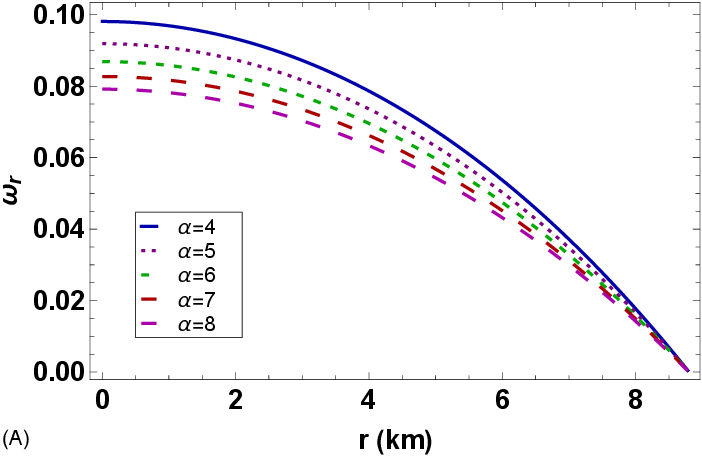}
\includegraphics[scale=.63]{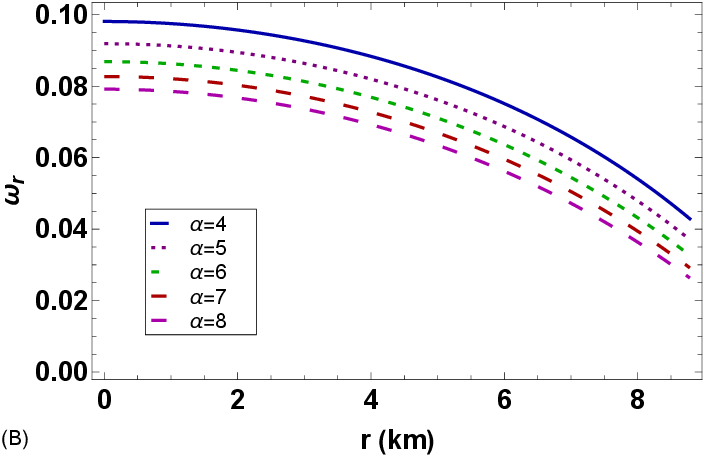}
\includegraphics[scale=.63]{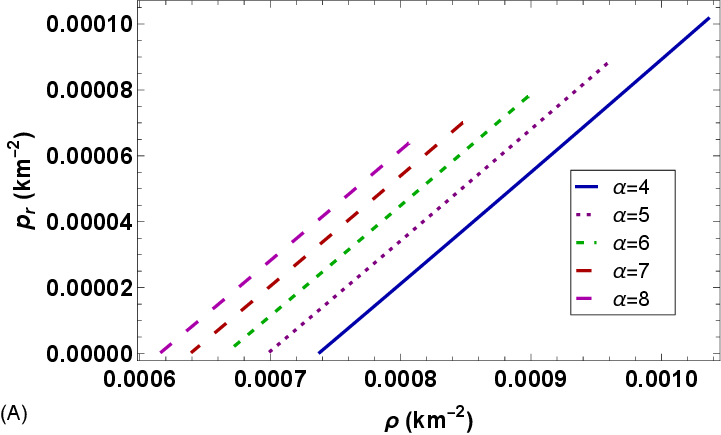}
\includegraphics[scale=.63]{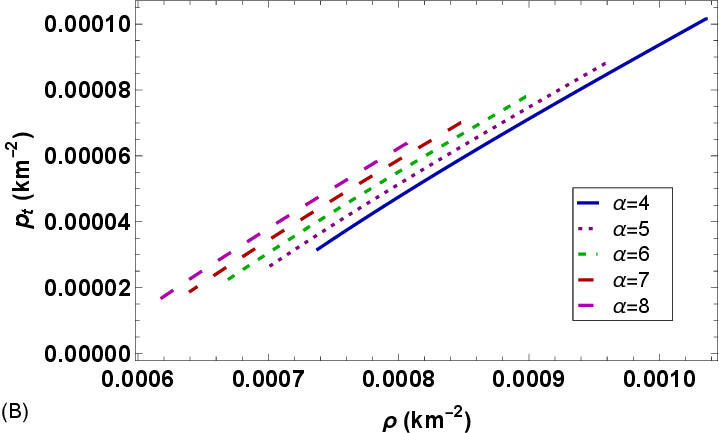}
\caption{Evolution of $\omega_r$ and $\omega_t$ with respect to $r$ as well as behavior of $p_r$ and $p_t$ with respect to $\rho$.}\label{eos}
\end{figure}

One should use an equation of state to link the microphysics to a physical star model and, effectively, to test the model physically, one should begin with an EoS. The EoS parameter is an important astrophysical instrument that may assist in our understanding of the fundamental aspects of the distribution of matter. We weigh the contributions of various elements, such as dark and baryonic slow-moving matter $(\omega = 0)$, ultra-relativistic matter $(\omega = 1/3)$, and dark energy (probably $\omega \sim -1)$), to determine the effectively combined $\omega (= p/\rho)$ of all matter in the universe. More attention is currently being paid to the equation of state of matter at ultra-high densities concerning the issue of the gravitational collapse of heavy star evolution in its last phase. As such, comprehensive $`\omega$' measurements will also provide information about the relative diversity of different materials. Regarding Zeldovich's requirement, any fluid sphere that is thought to be physically acceptable needs to have a pressure-to-density ratio that is positive and less than unity. This means that $\omega$ must always be continuous at the junction and lie between 0 and 1 \cite{l1962equation, zeldovich1971relativistic}.
Therefore, the EoS parameters, often represented by the two dimensionless values that can be used to characterize the connection between matter density and pressure, are
 $$\omega_r = \frac{p_r}{\rho},\omega_t = \frac{p_t}{\rho}.$$
Radial pressure and matter density are acknowledged to be quadratically related in the model by solving the field equations; nevertheless, the transverse pressure and matter density are still undetermined.
In Fig. \ref{eos}, we have illustrated their profiles to analyze the behavior of the equation of state parameter and observe that both $\omega_r$ and $\omega_t$ are monotonically decreasing functions of $r$; yet, they are both contained in the interval $0 < \omega_r, \omega_t < 1.$. The figure also illustrates how the density has affected the radial and transverse pressures. The important consequence drawn from this finding is that in an expanding universe, the energy density decreases more rapidly than the volume grows because of the wavelength of the radiation being red-shifted.
The stability of our proposed model is confirmed by subfigures (A) and (B) of Fig.~\ref{eos}, which we can frequently verify.

\section{Stability and feasibility analysis}\label{Sec5}
This section looks at a crucial mechanism called the stability mechanism.
However, it could be challenging to keep the model consistent when multiple variables are changing at once. In this regard, innovative techniques for evaluating the stability of a stellar model incorporate the cracking concept of Herrera (or the causality condition), relativistic adiabatic index, energy conditions (ECs), and static equilibrium through the TOV equation.

\subsection{Herrera's cracking concept}
\begin{figure}[H]
\centering
\includegraphics[scale=.63]{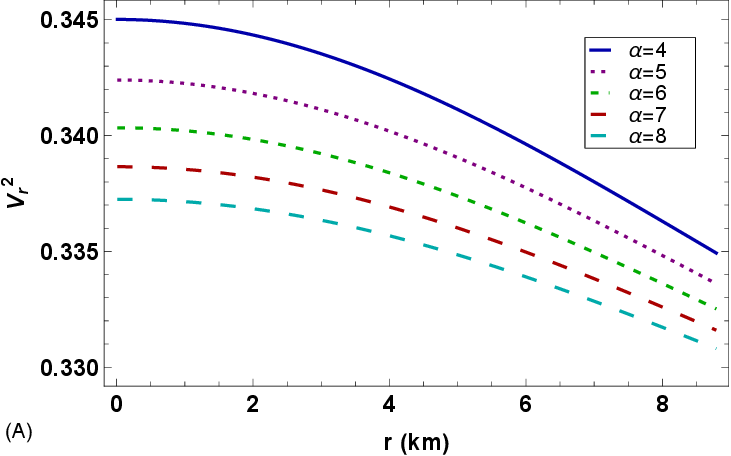}
\includegraphics[scale=.63]{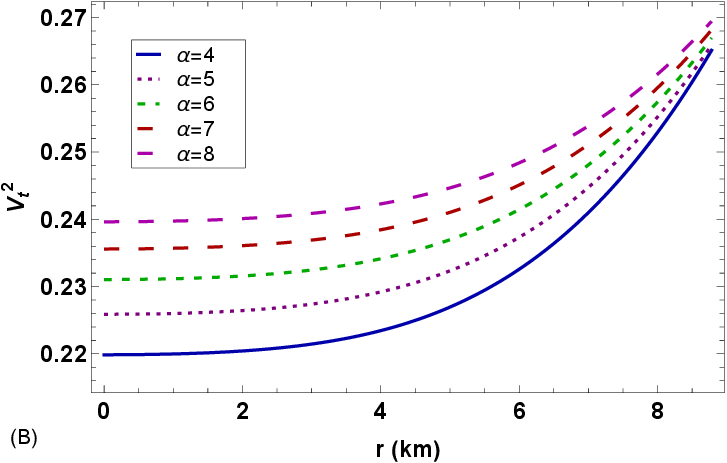}
\includegraphics[scale=.63]{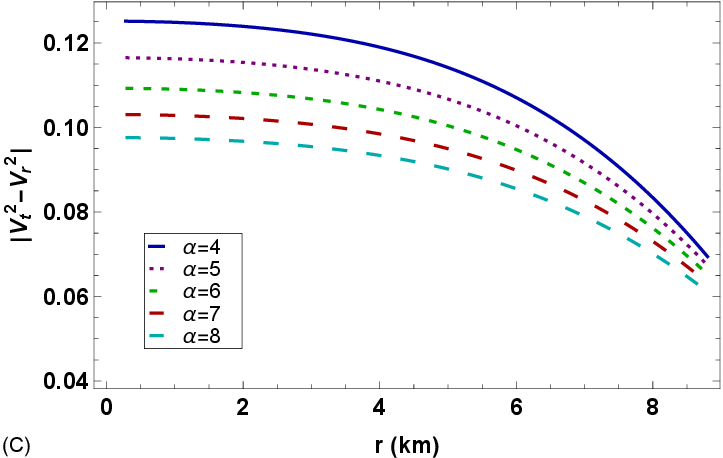}
\caption{(A, B) Square of the sound velocity components and (C) the stability factor $|V_t^2-V_r^2|$ have been plotted against radius $r$.}\label{speed}
\end{figure}

The stability of celestial bodies under radial perturbations resulting from anisotropic stresses within the fluid sphere has been established by the Herrera cracking method \cite{abreu2007sound}.
Few physical measurements exist for determining the size (absolute and/or relative) of the perturbation, or how small (or big) the perturbations should be when arbitrary and independent density and anisotropy disturbances are taken into account. Cracking could be caused by perturbations of varying magnitudes (and relative sizes $\delta\Delta/\delta\rho$). Variable perturbations may be a more effective way to produce cracking within a specific matter configuration. In this regard, the concept of cracking for self-gravitating systems through the use of ideal fluid and anisotropic material distributions was introduced by Herrera and his associates in their writings \cite{herrera1992cracking, di1994tidal, di1997cracking}. This method helps to identify potentially unstable anisotropic matter structures. Such a breakthrough idea was devised to explain how fluid distributions change instantly as they exit their equilibrium phases due to the existence of non-vanishing radial forces. This strategy can be simply explained as follows:
\begin{eqnarray}
\frac{\delta\Delta}{\delta\rho} \sim \frac{\delta (p_t - p_r)}{\delta\rho} \sim \Big(\frac{\delta p_t}{\delta\rho} - \frac{\delta p_r}{\delta\rho}\Big) \sim \big(V_t^2 - V_r^2\big),
\end{eqnarray}
where $V_r = \sqrt{\frac{dp_r}{d\rho}}$ and $V_t = \sqrt{\frac{dp_t}{d\rho}}$ represent the radial and tangential sound speeds, respectively. In any case, describing the celestial interior, the subliminal sound speed of the pressure waves must be less than the speed of light i.e. $0 \leq {V_r}^2 \leq c^2$ and $0 \leq {V_t}^2 \leq c^2$, (here, $c = 1$). The major axes of a substance are the radial and transverse directions in which pressure waves propagate while working with an anisotropic fluid. $V_r$ and $V_t$ indicate the speed of the subliminal sound in these directions. The truth is that the sound velocity is determined by the slope of the $p_r(\rho)$ and $p_t(\rho)$ functions. 
Now, it is important to verify the causality requirement, which states that the velocity of sound within the compact object must always be less than unity i.e. $0 \leq V_r^2 \leq 1 $ and $0 \leq V_t^2 \leq 1$ \cite{karmakar2023charged}, to develop a physically acceptable model.
Consequently, 
\begin{eqnarray}\label{crack}
|V_t^2 - V_r^2| \leq 1.
\end{eqnarray}
Additionally, it is possible to express equation (\ref{crack}) explicitly as
\cite{jasim2021anisotropic, gedela2021new}:
\begin{equation}\label{crack1}
|V_t^2 - V_r^2| \leq 1 =\left\{\begin{array}{ll}
                    -1 \leq V_t^2 - V_r^2 \leq 0,  & \qquad {\rm Potentially ~ stable}, \\
                   0 < V_t^2 - V_r^2 \leq 1, & \qquad {\rm Potentially ~ unstable}.
                  \end{array}
\right.
\end{equation}
In the above-mentioned extreme matter configurations, ($p_t \neq 0, p_r = 0$) are always theoretically stable for cracking, but ($p_t = 0, p_r \neq 0$) become possibly unstable. Furthermore, the magnitude of anisotropy perturbations should always be less than the size of density perturbations for physically plausible models, i.e.
$$|V_t^2 - V_r^2| \leq 1 \implies \delta\Delta \leq \delta\rho.$$
Potentially unstable models result from these perturbations when $\delta\Delta > \delta\rho$.\\
Ratanpal \cite{ratanpal2020cracking} shortly deduced a simplified method for verifying the cracking condition for a spherically static symmetric spacetime with decreasing matter density with r in terms of the gradient of anisotropy with respect to r. Potentially stable regions are those where $\frac{\delta\Delta}{\delta r} \geq 0$, while potentially unstable regions are those where $\frac{\delta\Delta}{\delta r} < 0$.\\
For different values of $\alpha$, the profiles of $V_r^2$ and $V_t^2$ are shown in the upper left and right panels of Fig. \ref{speed}. These profiles definitively show that both sound velocities lie within the predicted range $[0,1]$, supporting the notion that our model meets the causality requirement.

Inside the star, where the transverse velocity is less than the radial velocity of sound, the expectation of stability was studied. As we can see from the lower panel of Fig. \ref{speed}, for $\alpha = 4, 5, 6, 7$, and $8$ in $[0, \mathfrak{R}]$, the radial speed of sound is faster than the tangential speed of sound throughout the star, i.e., $|V_t^2 - V_r^2| \leq 1$ which shows that there is no crack in the cosmic core \cite{andreasson2009sharp}. Thus, Fig.~\ref{speed} illustrates that our current polytropic star model satisfies Herrera's cracking concept as well as the causality requirements, pointing out that it is theoretically consistent.

\subsection{Relativistic adiabatic index}
\begin{figure}[H]
\centering
\includegraphics[scale=.63]{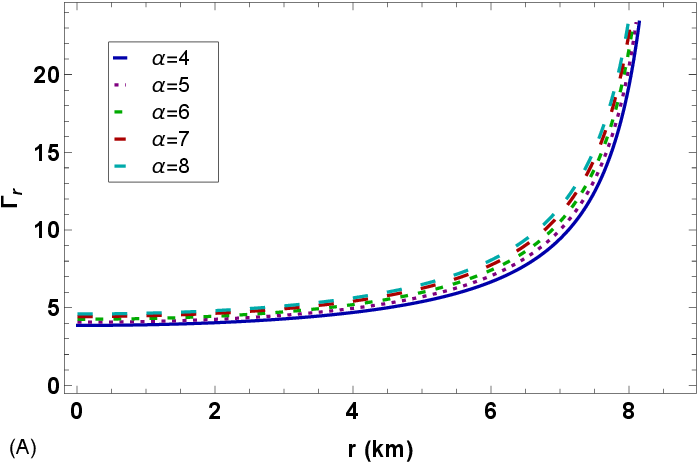}
\includegraphics[scale=.64]{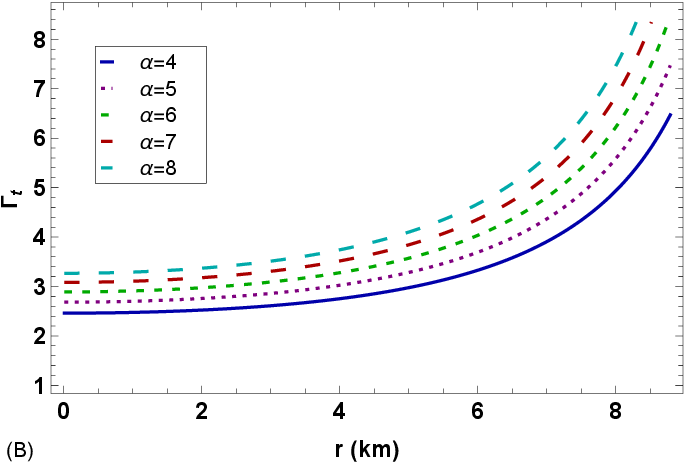}
\caption{Behavior of the adiabatic index $\Gamma_r$ and $\Gamma_t$ as a function of the radial distance $r$.}\label{gama}
\end{figure}

The ratio of two specific heats, called the adiabatic index $\Gamma$,
reflects the stiffness of the EoS for a specific density profile. The stability of a relativistic and non-relativistic fluid sphere can be studied with it.
The adiabatic index expressions are as follows when pressure anisotropy is present:
\begin{eqnarray}
\Gamma_r= \frac{\rho + p_r}{p_r}V_r^2,
\end{eqnarray} 
and 
\begin{eqnarray}
\Gamma_t= \frac{\rho + p_t}{p_t}V_t^2,
\end{eqnarray} 
Chandrasekhar \cite{chandrasekhar1964dynamical} proposed a method to investigate dynamical stability based on the variational method against an infinitesimal radial adiabatic perturbation. This methodology led to the discovery of a crucial relationship for the adiabatic index \cite{chandrasekhar1964dynamical, merafina1989systems}. Regarding this, several researchers have focused on studying the dynamical stability of star configurations \cite{Heint, Hilleb, Bombaci}. The stability criterion for a relativistic compact object is specified by $\Gamma > 4/3$ in the presence of a positive and growing anisotropy factor $\Delta = p_t - p_r > 0$, since a positive anisotropy factor can delay the manifestation of an instability \cite{heintzmann1975neutron}.
Radiation and anisotropy were recently included in this concept by Moustakidis \cite{moustakidis2017stability}. In the relativistic scenario, the above requirement changes for an isotropic sphere because of the influence of regeneration pressure, which makes the sphere more unstable. However, additional complexities develop for general relativistic anisotropic spheres, since the anisotropy determines the stability of the star system.
We can observe that $\Gamma_t$ and $\Gamma_r$ are both monotonically increasing functions of `$r$' and the adiabatic index is greater than $4/3$ inside the stellar interior for our proposed model, as shown in Fig. \ref{gama}. 
Therefore, from the perspective of the relativistic adiabatic index, our model is consistent.
A numerical comparison of the adiabatic index $\Gamma_r$ and $\Gamma_t$ at $r=0$ is given for different values of the coupling parameter $\alpha$ in Table \ref{table4}.
\begin{table}[h]
\begin{center}
\caption{\label{table4} Comparative study of the adiabatic index $\Gamma_r$ and $\Gamma_t$ at $r=0$.}
\begin{tabular}{|c|c|c|c|}
\hline
$\alpha$   &   $\Gamma_r$  &  $\Gamma_t$ \\
\hline\hline
4  & 3.86182 & 2.46089 \\
5  & 4.06883 & 2.68418 \\
6  & 4.25826 & 2.89084 \\
7  & 4.43374 & 3.08432 \\
8  & 4.59779 & 3.26700 \\
\hline
\end{tabular}
\end{center}
\end{table}
\~~~~\

\subsection{Energy Conditions}
It could be somewhat difficult to characterize the concrete form of the energy-momentum tensor, even though the elements that make up the matter distribution are known. Although one has certain theories about how the matter would behave in extreme conditions of density and pressure, it is permissible  to assume certain inequalities, called energy conditions, to verify the behavior of the energy-momentum tensor everywhere within the star \cite{bhar2019compact}. The distribution of mass, momentum, and stress resulting from the presence of matter and any non-gravitational fields is described by the energy-momentum tensor $T_{ij}$, which is derived from general relativity. On the other hand, neither the acceptable non-gravitational fields nor the state of matter in the space-time model are directly addressed by the Einstein field equations \cite{biswas2019strange}. Energy conditions are the terms used to describe how matter is distributed in space-time as measured by an observer. Fundamentally, in GR, the energy conditions allow for various non-gravitational fields and all states of matter, while also justifying the physically feasible solutions. We must determine whether our model satisfies all of the energy conditions to fully investigate the physical characteristics of an anisotropic strange star. It follows that matter should flow along the null or time-like world-line if these conditions are positive \cite{biswas2020anisotropic}. An important aspect of this work is the energy conditions, which need to be non-negative across the stellar medium and satisfied for every internal fluid sphere \cite{karmakar2024celestial}. In other words, for pressures and density to seem physically appropriate, they must be restricted to some extent. When the energy conditions (ECs) are applied to the substance parameters, one can can distinguish between a regular substance and an atypical fluid. The following energy conditions must be met for the current model to have any physical significance in this method:

\begin{itemize}
\item Strong Energy Condition (SEC):~ $\rho + p_r + 2p_t  \geq 0$\\

\item Weak Energy Condition (WEC):~ $\rho + p_r \geq 0, ~~~ \rho + p_t  \geq 0$\\

\item Null Energy Condition (NEC):~ $\rho  \geq 0$ \\

\item Dominant Energy Condition (DEC):~ $\rho - p_r  \geq 0,~~~\rho - p_t \geq 0$ \\ 

\item Trace Energy Condition (TEC):~ $\rho - p_r  - 2p_t \geq 0$ \\ 
\end{itemize}

\begin{figure}[H]
\centering
\includegraphics[scale=.6]{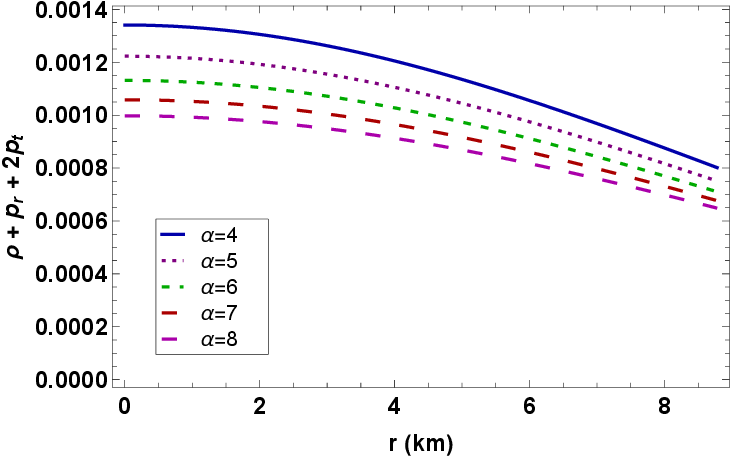}
\includegraphics[scale=.6]{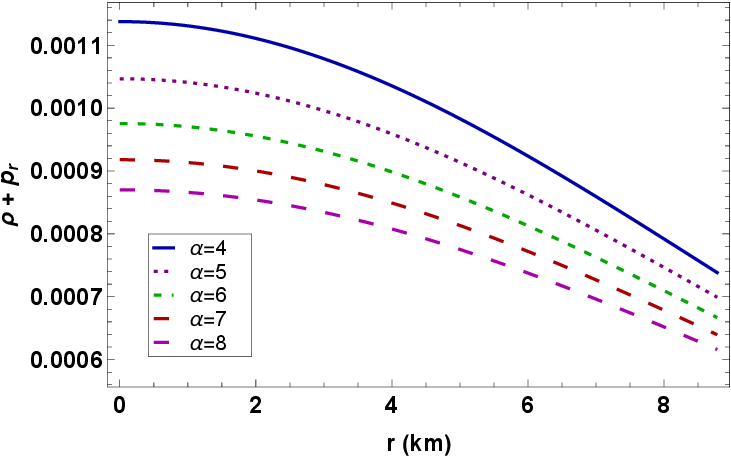}\\[1\baselineskip]
\includegraphics[scale=.6]{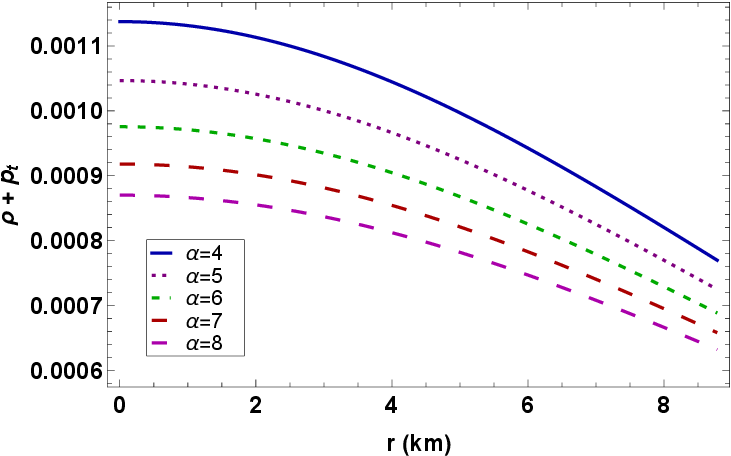}
\includegraphics[scale=.6]{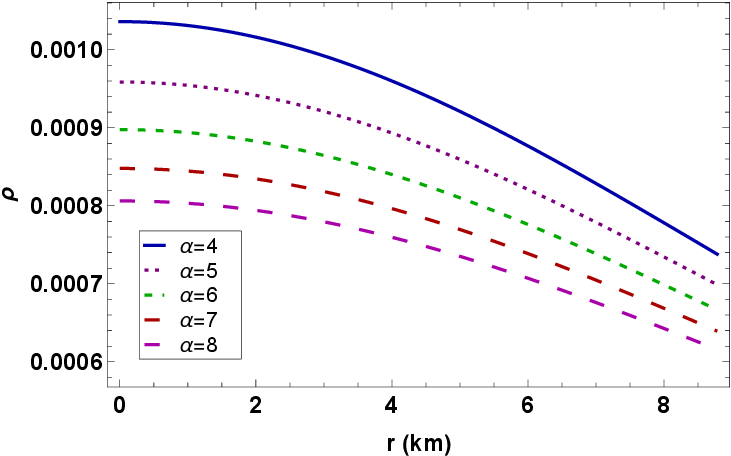}\\[1\baselineskip]
\includegraphics[scale=.6]{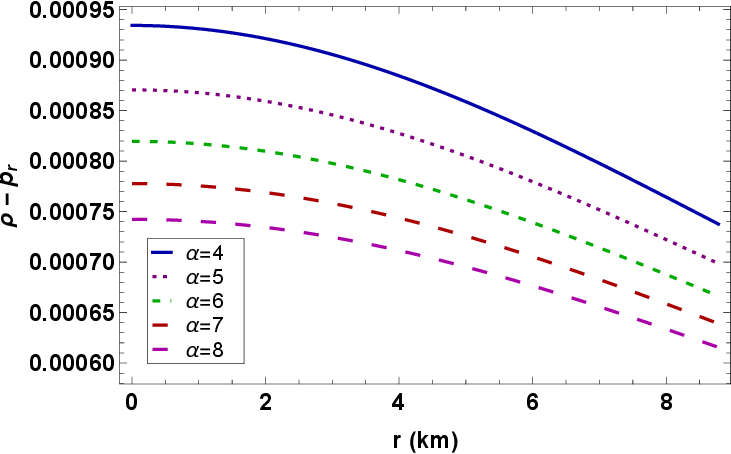}
\includegraphics[scale=.6]{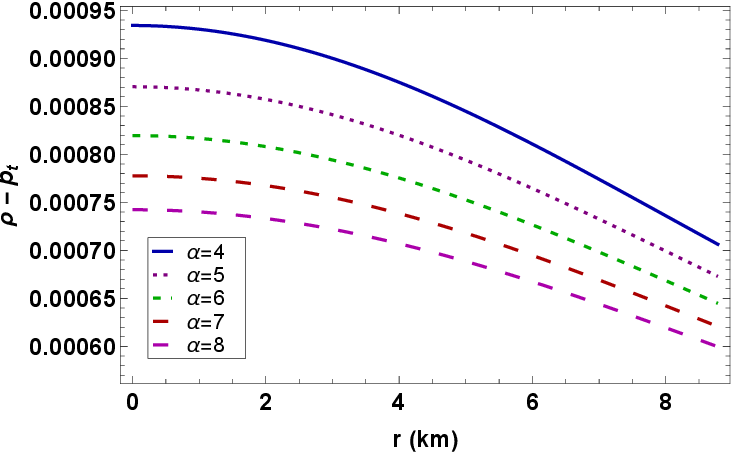}\\[1\baselineskip]
\includegraphics[scale=.6]{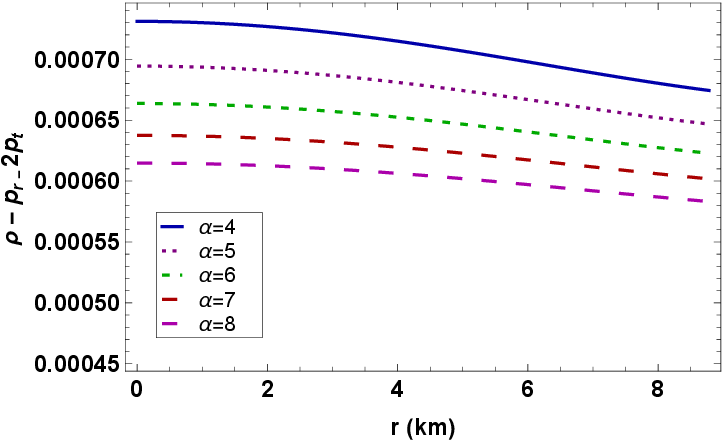}
\caption{Variation of the ECs with respect to the radial coordinate $r$.}\label{Ener}
\end{figure}

The profiles of the previously indicated energy criteria are fully satisfied, as we can see from their graphical representations in Fig.~\ref{Ener}, suggesting that the proposed compact stellar model is physically stable. 

\section{Hydrostatic equilibrium}\label{Sec6}

\begin{figure}[H]
\centering
\includegraphics[scale=.62]{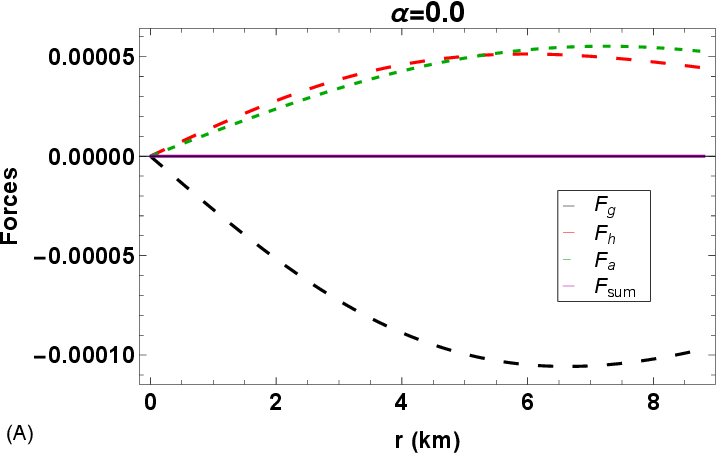}
\includegraphics[scale=.62]{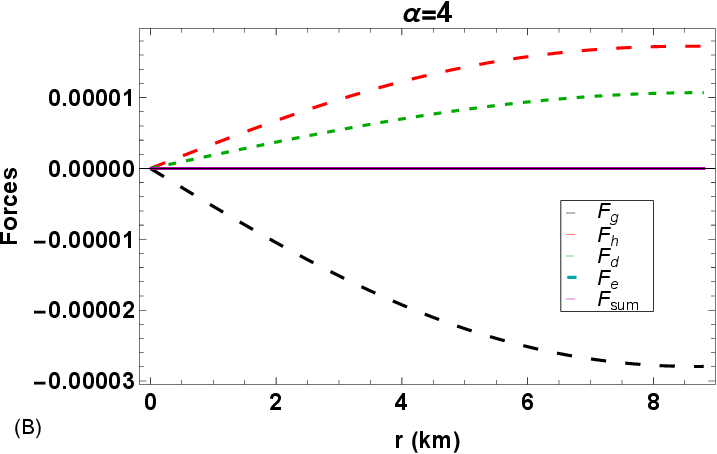}\\
\includegraphics[scale=.62]{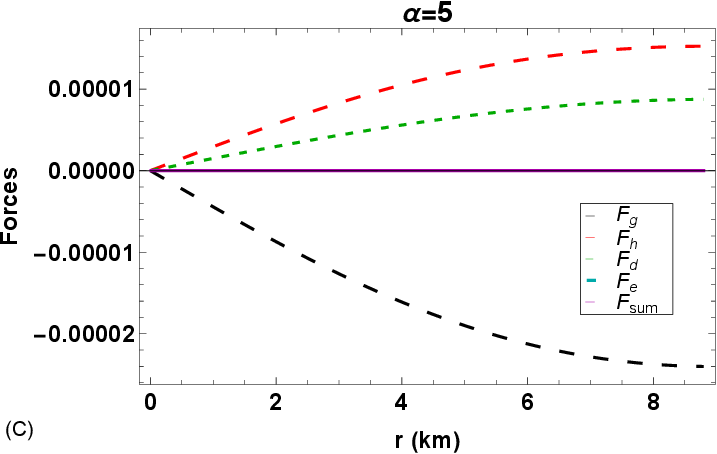}
\includegraphics[scale=.62]{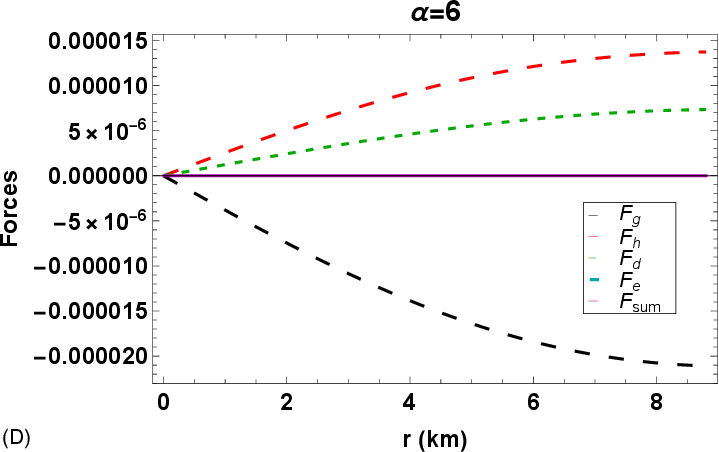}\\
\includegraphics[scale=.62]{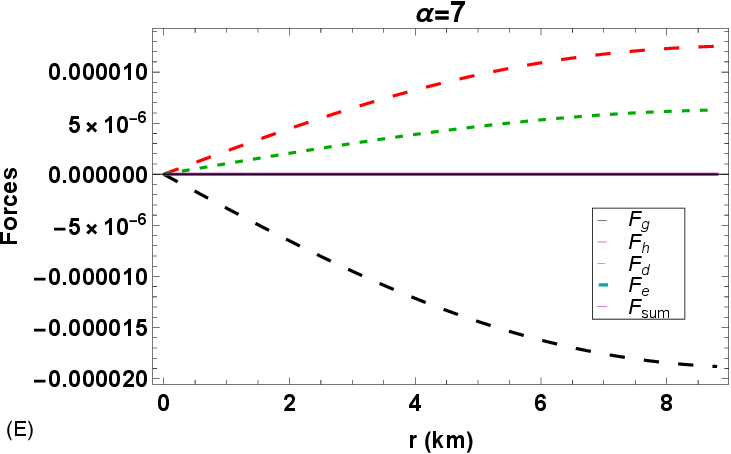}
\includegraphics[scale=.62]{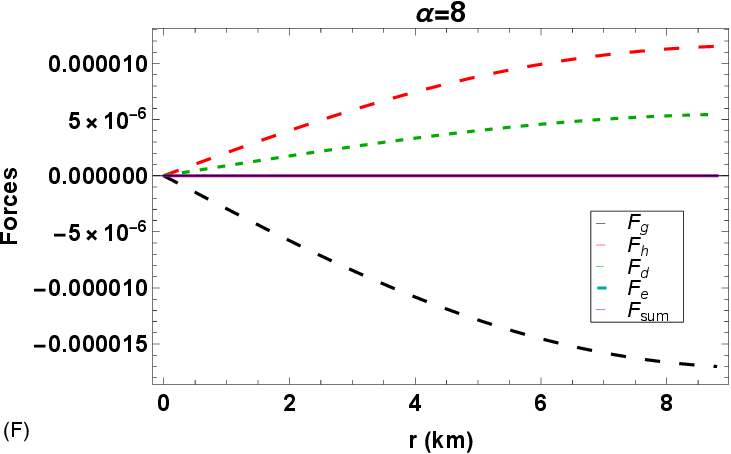}
\caption{Variations of forces with respect to $r$ for different values of $\alpha$.}\label{force}
\end{figure}

We must first conduct a comprehensive investigation of the stellar structure of compact stars to examine them. EoS describes the internal structure of any compact object and determines its stellar attributes. When the outward and repulsive forces created inside the stellar object balance the inward gravitational force so that the net force acting on the system is zero, the compact stellar system is said to be stable. Nevertheless, the system will become unstable in response to even a slight perturbation. An important feature of the provided physically realistic compact item is the hydrostatic equilibrium equation.\\
According to the known Einstein field equations for EGB gravity theory, the energy conservation for our stellar model is $\nabla^{i}T_{ij} = 0$. Consequently, the conservation equation of the energy-momentum tensor suggests the generalized Tolman-Oppenheimer-Volkoff (TOV) equation  \cite{tolman1939static, oppenheimer1939massive} as follows:  
\begin{equation}\label{tov1}
F_{sum}=F_g + F_h + F_a = 0,
\end{equation}
which can be utilized to evaluate the state of equilibrium equation for our compact star candidate under the combined action of several forces.
where $$F_g = -\frac{\nu'}{2}(\rho+p_r),$$ 
$$F_h = -\frac{dp_r}{dr},$$ 
$$F_a = \frac{3}{r}(p_t - p_r)$$
respectively, correspond to the gravitational force, the hydrodynamic force of ordinary matter, and the anisotropic force. 
For various values of $\alpha$, we present the profiles of the forces $F_g$, $F_h$, and $F_a$ along with $F_{sum}$ in Fig.\ref{force} to ensure the equilibrium state of the suggested stellar structure. In conclusion, the attainment of static equilibrium is made possible by the forces mentioned above.

\section{Concluding Remarks}\label{con}

In the present work, we have assumed a generalized Finch-Skea metric in the framework of 5$\mathcal{D}$ EGB gravity to construct a stellar model.
The internal structure of the proposed polytropic star is ultimately influenced by the GB term, which is constructed from quadratic contractions of the Riemann and Ricci tensors.
By considering the polytropic equation of state described in the relationship between the radial pressure and the normal baryonic matter density, we found the solutions of the metric coefficient $\nu(r)$ as well as the density, radial pressure, and tangential pressure. 
All physical requirements and the static stability requirement are met by this approach. Thus, it is possible to conclude that the polytropic star found in this study can comprehend a compact stellar region.
In our configured stellar object, we have considered the compact star EXO 1785-248 and then obtained the values of the free parameters $A$ and $\chi$ in Table \ref{table1} for different values of $\alpha$. To satisfy the regularity of the metric coefficients, we have drawn them graphically and observed that they always increase against $r$. On the other hand, to confirm the regularity of the fluid components (matter density, radial pressure, and tangential pressure), we have drawn them graphically and shown that they always cease to decrease with $r$. The gradients of the matter density and pressure components also decrease with $r$. In the stellar model, we have examined the requirements for a realistic star configuration. Additionally, the implications of pressure anisotropy and trace profile on compact celestial objects are examined. Further, we have analyzed the behaviors of the compactness factor and the mass function for the stellar object and shown that they are increasing in nature. Moreover, the inner and surface redshift functions have been shown graphically, and interestingly, their behavior is opposite. We have also discussed the nature of radial and tangential pressures as well as their equation of state parameters. We have also seen that the polytropic star model satisfies Herrera's cracking concept as well as the causality requirements, which points out that it is theoretically consistent. In addition, we have observed that the adiabatic index $\Gamma_r$ and $\Gamma_t$ both increase monotonically and that the adiabatic index is greater than $4/3$ inside the stellar model, so our model is consistent. Then, we have examined all the energy conditions, namely the NEC, WEC, SEC, DEC, and TEC for the polytropic star model, and these continuously show positive values throughout the stellar medium, indicating that our proposed model is attainable. Finally, we have presented all the forces (i.e. gravitational force, hydrodynamic force, and anisotropic force) from the TOV equation in the attainment of static equilibrium.\\ 
As a result, we can say that our findings might encourage researchers to look for polytropic stars of this kind, as the celestial structure we developed in this paper satisfies the required astrophysical requirements.


\section*{Author contributions}
\textbf{Akashdip Karmakar} performed original draft preparation, mathematical analysis, computer code design for data analysis, and numerical data analysis. \textbf{Ujjal Debnath} contributed to validation, methodology, writing - review, and editing.
\textbf{Pramit Rej} contributed to conceptualization, validation, investigation, overall supervision, writing - review \& editing of the project. All authors have read and approved the entire manuscript.

\section*{Acknowledgements} 
Pramit Rej is thankful to the Inter-University Centre for Astronomy and Astrophysics (IUCAA), Pune, Government of India, for providing a Visiting Associateship.

\section*{Declarations}
\textbf{Funding:} The authors did not receive any funding in the form of financial aid or grants from any institution or organization for the present research work.\par
\textbf{Data Availability Statement:} The results are obtained
through purely theoretical calculations and can be verified analytically;
thus, this manuscript has no associated data, or the data will not be deposited. \par
\textbf{Conflicts of Interest:} The authors declare that they have no known competing financial interests or personal relationships that could have appeared to influence the work reported in this paper.\par

\bibliographystyle{apsrev4-1}
\bibliography{poly_references}

\newpage

{\textbf{Appendix:}}
\begin{eqnarray}
f_1(r) &=& [512 \pi^3 \chi (9 + 8 \pi r^2 \chi) + 64 A^2 \pi (162 (\gamma + 8 \alpha \beta \pi) + 
      9 (8 + \beta (195 + 16 \beta)) \pi r^2  + 24 \pi r^2 (12 \gamma + 4 \alpha (103 + 8 \beta) \pi \nonumber\\&& + (567 + 
            142 \beta) \pi r^2) \chi  + 12160 \pi^3 r^6 \chi^2) + 192 A^3 \pi (288 \alpha (3 \gamma + 4 \alpha \beta \pi) + 
      3 (3 (75 + 16 \beta) \gamma + 32 \alpha (1 + \beta (59 + 4 \beta)) \pi) r^2 + \nonumber\\&&
      6 (65 + 2 \beta (379 + 62 \beta)) \pi r^4 + 16 \pi r^2 (48 \alpha \gamma + (93 \gamma + 
            4 \alpha (415 + 62 \beta) \pi) r^2 + 4 (421 + 149 \beta) \pi r^4)\chi + 24320 \pi^3 r^8 \chi^2) + \nonumber\\&&
   2 A^{18} r^{28} (9 (144 \alpha \gamma^2 + 9 \gamma (97 \gamma + 16 \alpha (3 + 4 \beta) \pi) r^2 + 
         4 \pi (3 (383 + 386 \beta) \gamma + 32 \alpha (2 + \beta (7 + 4 \beta)) \pi) r^4 + \nonumber\\&&
         32 (190 + \beta (383 + 192 \beta)) \pi^2 r^6) + 768 \pi^2 r^4 (12 \alpha \gamma + 
         4 (36 \gamma + \alpha (5 + 4 \beta) \pi) r^2 + (383 + 382 \beta) \pi r^4) \chi \nonumber\\&& + 389120 \pi^4 r^{10} \chi^2) + 
   12 A^{17} r^{26} (3672 \alpha \gamma^2 + 9 \gamma (903 \gamma + 32 \alpha (38 + 51 \beta) \pi) r^2 + 
      24 \pi (3 (581 + 594 \beta) \gamma + 16 \alpha (17 + 60 \beta \nonumber\\&& + 34 \beta^2) \pi) r^4 + 
      96 (569 + 2 \beta (582 + 293 \beta)) \pi^2 r^6 + 128 \pi^2 r^4 (204 \alpha \gamma + (879 \gamma + 
            16 \alpha (22 + 17 \beta) \pi) r^2 + 4 (583 + 578 \beta) \pi r^4) \chi \nonumber\\&& + 
      389120 \pi^4 r^{10} \chi^2) + 96 A^5 (3 (4608 \alpha^3 \gamma \pi + 384 \alpha \gamma (3 \gamma + \alpha (43 + 12 \beta) \pi) r^2 + 
         2 (396 \gamma^2 + 9 \alpha (1235 + 704 \beta) \gamma \pi \nonumber\\&&  + 32 \alpha^2 \beta (489 + 
               88 \beta) \pi^2) r^4 + \pi (3 (3557 + 2630 \beta) \gamma + 16 \alpha (365 + 2 \beta (2820 + 653 \beta)) \pi) r^6 + 
         4 (2383 + 2 \beta (6625 \nonumber\\&& + 2063 \beta)) \pi^2 r^8) + 16 \pi^2 r^4 (2112 \alpha^2 \gamma  + 
         7836 \alpha \gamma r^2 + (6189 \gamma + 280 \alpha (209 + 58 \beta) \pi) r^4 + 272 (163 + 86 \beta) \pi r^6) \chi \nonumber\\&& + 
      661504 \pi^4 r^{12} \chi^2) + 24 A^{15} r^{20} (3 (6048 \alpha^2 \gamma^2 +  192 \alpha \gamma (132 \gamma + \alpha (-43 + 84 \beta) \pi) r^2 + (19971 \gamma^2 + 72 \alpha (913 + 1382 \beta) \gamma \pi \nonumber\\&& + 256 \alpha^2 \beta (-41 + 28 \beta) \pi^2) r^4 + 8 \pi (9 (1333 + 1431 \beta) \gamma + 16 \alpha (325 + 2 \beta (580 + 339 \beta)) \pi) r^6 \nonumber\\&& + 16 (7609 + 2 \beta (8151 + 4151 \beta)) \pi^2 r^8) + 64 \pi^2 r^4 (672 \alpha^2 \gamma + 8136 \alpha \gamma r^2 + (12453 \gamma + 40 \alpha (373 + 266 \beta) \pi) r^4 \nonumber\\&& + 544 (61 + 59 \beta) \pi r^6) \chi + 2646016 \pi^4 r^{12} \chi^2) + 20 A^{19} r^{32} (9 \gamma + 8 \pi r^2 (3 + 3 \beta + 8 \pi r^2 \chi))^2 \nonumber\\&& + A^{20} r^{34} (9 \gamma + 8 \pi r^2 (3 + 3 \beta + 8 \pi r^2 \chi))^2 + 256 A \pi^2 (3 \beta (9 + 16 \pi r^2 \chi) + 8 \pi \chi (18 \alpha + 5 r^2 (9 + 8 \pi r^2 \chi))) \nonumber\\&& + 24 A^4 (864 \gamma^2 r^2 + 9 \gamma \pi r^4 (3623 + 1696 \beta + 9648 \pi r^2 \chi) + 8 \pi^2 r^6 (2985 + 3 \beta (7502 + 1809 \beta) + 1360 (83 + 37 \beta) \pi r^2 \chi \nonumber\\&& + 103360 \pi^2 r^4 \chi^2)  + 768 \alpha^2 \pi (4 \beta (29 + 2 \beta) \pi r^2 + 
         \gamma (45 + 16 \pi r^2 \chi)) + 96 \alpha \pi r^2 (\gamma (663 + 144 \beta + 848 \pi r^2 \chi) \nonumber\\&& + \pi r^2 (112 + 2913 \beta + 
            424 \beta^2 + 80 (139 + 30 \beta) \pi r^2 \chi))) + 24 A^6 r^2 (9216 \alpha^3 (13 + 8 \beta) \gamma \pi + 
      48708 \gamma^2 r^4 \nonumber\\&& + 81 (4569 + 4510 \beta) \gamma \pi r^6 + 48 (8005 + 4 \beta (8871 + 3298 \beta)) \pi^2 r^8  + 
      4352 \pi^2 r^8 (291 \gamma + (1595 + 958 \beta) \pi r^2) \chi \nonumber\\&& + 6615040 \pi^4 r^{12} \chi^2 + 
      192 \alpha^2 (432 \gamma^2 + 4 \beta (1147 + 442 \beta) \pi^2 r^4 + \gamma \pi r^2 (1725 + 2520 \beta + 3536 \pi r^2 \chi)) + 
      4 \alpha r^2 (30240 \gamma^2 \nonumber\\&& + 9 \gamma \pi r^2 (19095 + 21432 \beta + 52928 \pi r^2 \chi) + 8 \pi^2 r^4 (8814 + 3 \beta (30821 + 9924 \beta) + 
            1456 (209 + 70 \beta) \pi r^2 \chi))) \nonumber\\&& +  3 A^{16} r^{22} (143829 \gamma^2 r^4  + 24 \gamma \pi r^6 (29913 + 31230 \beta + 81344 \pi r^2 \chi) + 128 \pi^2 r^8 (7215 + 186 \beta (81 + 41 \beta) \nonumber\\&& + 68 (595 + 584 \beta) \pi r^2 \chi + 51680 \pi^2 r^4 \chi^2) + 
      384 \alpha^2 (27 \gamma^2 + 16 \beta (-3 + 2 \beta) \pi^2 r^4 + 4 \gamma \pi r^2 (-9 + 18 \beta + 16 \pi r^2 \chi))  + \nonumber\\&&
      192 \alpha r^2 (621 \gamma^2 + 2 \gamma \pi r^2 (879 + 1233 \beta + 2176 \pi r^2 \chi) + 4 \pi^2 r^4 (268 + 951 \beta + 544 \beta^2 + 
            16 (121 + 90 \beta) \pi r^2 \chi))) \nonumber\\&& + 2 A^{10} r^6 (3981312 \alpha^4 \gamma^2 + 27648 \alpha^3 \gamma r^2 (726 \gamma + (-2341 + 
            900 \beta) \pi r^2) + 1728 \alpha^2 (20637 \gamma^2 r^4 + 32 \beta (-467 + 686 \beta) \pi^2 r^8 + \nonumber\\&&
         112 \gamma \pi r^6 (-353 + 465 \beta + 392 \pi r^2 \chi)) + 144 \alpha r^6 (194373 \gamma^2 + 
         3 \gamma \pi r^2 (83763 + 249516 \beta + 427328 \pi r^2 \chi) \nonumber\\&& + 88 \pi^2 r^4 (2730 + 3 \beta (4065 + 2428 \beta) + 
            2080 (17 + 9 \beta) \pi r^2 \chi)) + r^8 (8268561 \gamma^2 + 396 \gamma \pi r^2 (83589 + 108726 \beta + \nonumber\\&&
            282880 \pi r^2 \chi) + 9152 \pi^2 r^4 (4527 + 9 \beta (1357 + 680 \beta) + 408 (95 + 78 \beta) \pi r^2 \chi + 
            41344 \pi^2 r^4 \chi^2))) + 3 A^{12} r^{10} (110592 \alpha^4 \gamma^2 \nonumber\\&& +  18432 \alpha^3 \gamma r^2 (159 \gamma + (-575 + 
            208 \beta) \pi r^2) +  1152 \alpha^2 (8595 \gamma^2 r^4 +  160 \beta (-67 + 58 \beta) \pi^2 r^8 + 
         128 \gamma \pi r^6 (-117 + 171 \beta \nonumber\\&& + 145 \pi r^2 \chi)) + 128 \alpha r^6 (86805 \gamma^2 + 
         66 (1768 + \beta (6761 + 4216 \beta)) \pi^2 r^4  + 2912 (397 + 242 \beta) \pi^3 r^6 \chi + 6 \gamma \pi r^2 (24561 \nonumber\\&& + 54945 \beta + 
            92752 \pi r^2 \chi)) +  r^8 (4131189 \gamma^2 + 264 \gamma \pi r^2 (65763 + 79446 \beta + 201760 \pi r^2 \chi) + 
         1664 \pi^2 r^4 (13002 \nonumber\\&& + 3 \beta (10376 + 5335 \beta) + 272 (334 + 299 \beta) \pi r^2 \chi +  103360 \pi^2 r^4 \chi^2))) + 
   4 A^9 r^4 (2654208 \alpha^4 \gamma^2 + 27648 \alpha^3 \gamma r^2 (366 \gamma \nonumber\\&& + (-1103 +  462 \beta) \pi r^2) + 3456 \alpha^2 (4617 \gamma^2 r^4 + 8 \beta (-193 + 656 \beta) \pi^2 r^8 + 2 \gamma \pi r^6 (-4317 + 6030 \beta + 5248 \pi r^2 \chi))]
\end{eqnarray}

\newpage
\begin{eqnarray}
f_2(r) &=& [24 \alpha r^6 (488727 \gamma^2 + 12 \gamma \pi] r^2 (53001 + 162765 \beta + 288112 \pi r^2 \chi) +  176 \pi^2 r^4 (3471 + 6 \beta (2954 + 1637 \beta) \nonumber\\&& + 520 (103 + 50 \beta) \pi r^2 \chi)) + r^8 (3324321 \gamma^2 + 216 \gamma \pi r^2 (63179 + 83078 \beta + 223184 \pi r^2 \chi) + 416 \pi^2 r^4 (40815 \nonumber\\&& + 18 \beta (6647 + 3219 \beta) + 8160 (49 + 38 \beta) \pi] r^2 \chi + 413440 \pi^2 r^4 \chi^2))) + 4 A^{11} r^8 (663552 \alpha^4 \gamma^2 + 13824 \alpha^3 \gamma r^2 (435 \gamma \nonumber\\&& + 2 (-749 + 276 \beta) \pi r^2) + 1728 \alpha^2 (7947 \gamma^2 r^4 + 
         16 \beta (-501 + 524 \beta) \pi^2 r^8 + 8 \gamma \pi r^6 (-1847 + 2499 \beta + 2096 \pi r^2 \chi)) \nonumber\\&& + 
      96 \alpha r^6 (129384 \gamma^2 + 9 \gamma \pi r^2 (20823 + 54594 \beta + 92224 \pi r^2 \chi) + 176 \pi^2 r^4 (936 + 9 \beta (423 + 262 \beta) + 
            26 (403 + 230 \beta) \pi r^2 \chi)) \nonumber\\&& + r^8 (4020921 \gamma^2 + 1584 \gamma \pi r^2 (10302 + 12975 \beta + 33176 \pi r^2 \chi) + 
         416 \pi^2 r^4 (99 (495 + 2 \beta (625 + 319 \beta)) \nonumber\\&& + 3264 (115 + 99 \beta) \pi r^2 \chi + 413440 \pi^2 r^4 \chi^2))) + 
   12 A^{14} r^{16} (4608 \alpha^3 \gamma (3 \gamma + (-11 + 4 \beta) \pi r^2) + 96 \alpha^2 (2457 \gamma^2 r^2 \nonumber\\&& + 16 \beta (-249 + 178 \beta) \pi^2 r^6 + 4 \gamma \pi r^4 (-897 + 1620 \beta + 1424 \pi r^2 \chi)) + 8 \alpha r^4 (67662 \gamma^2 \nonumber\\&& + 3 \gamma \pi r^2 (51975 + 87408 \beta + 
            150464 \pi r^2 \chi) + 8 \pi^2 r^4 (13026 + 3 \beta (15663 + 9404 \beta) + 560 (191 + 130 \beta) \pi r^2 \chi)) \nonumber\\&& + 
      r^6 (310149 \gamma^2 + 6 \gamma \pi r^2 (237849 + 264318 \beta + 675584 \pi r^2 \chi) + 32 \pi^2 r^4 (55965 + 168 \beta (736 + 377 \beta) \nonumber\\&& 
      + 1360 (251 + 238 \beta) \pi r^2 \chi + 413440 \pi^2 r^4 \chi^2))) + 48 A^7 r^2 (2304 alpha^3 \gamma (48 \gamma + (-115 + 
            104 \beta) \pi r^2) \nonumber\\&& + 96 \alpha^2 (2808 \gamma^2 r^2 + 20 \beta (303 + 268 \beta) \pi^2 r^6 + \gamma \pi r^4 (-1947 + 9936 \beta + 10720 \pi r^2 \chi)) + 12 \alpha r^4 (20196 \gamma^2 \nonumber\\&& + \gamma \pi r^2 (52941 + 99774 \beta + 208064 \pi r^2 \chi)  + 16 \pi^2 r^4 (2041 + 15828 \beta + 6502 \beta^2 + 364 (139 + 54 \beta) \pi r^2 \chi)) +  r^6 (77301 \gamma^2 \nonumber\\&&  + 12 \gamma \pi r^2 (34905 + 40941 \beta + 
            125608 \pi r^2 \chi) + 8 \pi^2 r^4 (60081 + 42 \beta (5331 + 2243 \beta) + 2176 (389 + 259 \beta) \pi r^2 \chi \nonumber\\&& + 
            826880 \pi^2 r^4 \chi^2))) + 24 A^{13} r^14 (2304 \alpha^3 \gamma (33 \gamma + 2 (-61 + 22 \beta) \pi r^2) + 384 \alpha^2 (1215 \gamma^2 r^2 + 
         2 \beta (-883 + 680 \beta) \pi^2 r^6 \nonumber\\&& + \gamma \pi r^4 (-1941 + 3150 \beta + 2720 \pi r^2 \chi)) + 
      24 \alpha r^4 (29457 \gamma^2 + \gamma \pi r^2 (58767 + 112788 \beta + 191840 \pi r^2 \chi) + 8 \pi^2 r^4 (5291  \nonumber\\&& + 19488 \beta + 11990 \beta^2 + 
            728 (65 + 42 \beta) \pi r^2 \chi)) + r^6 (317115 \gamma^2 + 36 \gamma \pi r^2 (38699 + 44770 \beta + 113776 \pi r^2 \chi)  \nonumber\\&& + 
         16 \pi^2 r^4 (39 (2785 + 6382 \beta + 3282 \beta^2) + 1088 (647 + 598 \beta) \pi r^2 \chi + 826880 \pi^2 r^4 \chi^2))) + 
   3 A^8 r^2 (1769472 \alpha^4 \gamma^2  \nonumber\\&& + 18432 \alpha^3 \gamma r^2 (408 \gamma + (-1157 + 584 \beta) \pi r^2) + 4608 \alpha^2 (2709 \gamma^2 r^4 + 4 \beta (243 + 910 \beta) \pi^2 r^8 \nonumber\\&& + \gamma \pi r^6 (-4369 + 7770 \beta + 7280 \pi r^2 \chi)) + 64 \alpha r^6 (147555 \gamma^2 + 6 \gamma \pi r^2 (39777 + 105885 \beta + 199232 \pi r^2 \chi) \nonumber\\&& + 44 \pi^2 r^4 (4524 + 3 \beta (9231 + 4528 \beta) + 208 (415 + 182 \beta) \pi] r^2 \chi)) + r^8 (2725947 \gamma^2 + 24 \gamma \pi r^2 (509739 + 652266 \beta \nonumber\\&& + 1846208 \pi r^2 xi) + 1664 \pi^2 r^4 (8925 + 13314 \beta^2 + 136 \pi r^2 \chi (757 + 760 \pi r^2 \chi) + 34 \beta (855 + 2192 \pi r^2 \chi))))],
\end{eqnarray}

\end{document}